\documentclass[10pt]{article}
\usepackage{amsmath}                    
\usepackage{amssymb}                    
\usepackage{amsthm}                     
\usepackage{amsfonts}                   
\usepackage{graphicx}
\usepackage{subfigure}
\usepackage{graphics}
\usepackage{verbatim}
\newtheorem{rem}{Remark}[section]

\newtheorem{remark}{Remark}[section]
\newtheorem{theorem}{Theorem}[section]

\newtheorem{prop}{Proposition}[section]
\newtheorem{lemma}{Lemma}[section]
\newtheorem{result}{Result}[section]
\def \IR{\mathbb R}

\def \IE{\mathbb E}

\newcommand{\E}{{\mathbb E}}

\newcommand{\bX}{{\mathbb X}}
\newcommand{\CP}{{\mathcal P}}
\newcommand{\cS}{{\mathcal S}}

\newcommand{\CC}{{\mathcal C}}
\newcommand{\cO}{{\mathcal O}}
\newcommand{\CH}{{\mathcal H}}
\newcommand{\LL}{{\mathcal L}}

\newcommand{\cK}{{\mathcal K}}

\newcommand{\T}{{\mathbb T}}
\newcommand{\R}{{\mathbb R  }}

\newcommand{\bbR}{\mathbb{R}}
\newcommand{\cA}{\mathcal A}
\newcommand{\cL}{\mathcal L}

\newcommand{\eps}{\epsilon}

\title{Homogenization for Inertial Particles in a Random Flow}
\author{G. A. Pavliotis\thanks{Department of Mathematics, Imperial College
London, London SW7~2AZ, UK \newline(g.pavliotis@imperial.ac.uk) } \and A. M.
Stuart\thanks{Mathematics Institute, Warwick University, Coventry, CV4~7AL, UK
\newline (stuart@maths.warwick.ac.uk).} \and K.C. Zygalakis\thanks{Mathematics Institute,
Warwick University, Coventry, CV4~7AL, UK \newline (zygk@maths.warwick.ac.uk).} }

\begin{document}
\maketitle

\begin{abstract}
We study the problem of homogenization for inertial particles moving in a  time dependent
random velocity field and subject to molecular diffusion. We show that, under appropriate
assumptions on the velocity field, the large--scale, long--time behavior of the inertial
particles is governed by an effective diffusion equation for the position variable alone.
This is achieved by the use of a formal multiple scales expansion in the scale
parameter. The expansion relies on the hypoellipticity of the underlying diffusion. An
expression for the diffusivity tensor is found and various of its properties are studied.
The results of the formal multiscale analysis are justified rigorously by the use of the
martingale central limit theorem. Our theoretical findings are supported by numerical
investigations where we study the parametric dependence of the effective diffusivity on
the various non--dimensional parameters of the problem.
\end{abstract}

\section{Introduction}
\label{sec:intro}
Inertial particles play an important role in various applications in science and
engineering. Examples include planet formation, particle aggregation in rotating flows,
atmosphere/ocean science (in particular rain initiation \cite{falko_clouds,
shaw_clouds}), chemical engineering. The prominent role that inertial particles play in
various scientific and industrial applications has triggered many theoretical
investigations, see for example \cite{Max87, MaxRil83} and the references therein.

The starting point for many theoretical investigations concerning inertial particles is
Stokes' law which says that the force exerted by the fluid on the particle is
proportional to the difference between the background  fluid velocity and the particle
velocity:
\begin{equation}\label{e:stokes_intro}
F_{\tiny{s}}(t) \propto \Bigl( v(x(t),t) - \dot{x} \Bigr).
\end{equation}
Various extensions of this basic model have been considered in the literature, in
particular by Maxey and collaborators \cite{Max87,maxey_4,maxey_1,MaxRil83,maxey_5,
maxey_3}. In this work we will restrict ourselves to the analysis of particles subject to
a force of the form \eqref{e:stokes_intro}, together with additional molecular
bombardment.

In principle the fluid velocity $v(x,t)$ satisfies either the Euler or the Navier Stokes
equations and it is obtained through direct numerical simulations (DNS). The solution of
a Newtonian particle governed by force law \eqref{e:stokes_intro} coupled to either the
Euler or Navier--Stokes equations is analytically difficult to study and computationally
expensive. It is hence useful to consider $v(x,t)$ in \eqref{e:stokes_intro} to be  a
given random field $v(x,t)$ which mimics some of the features of velocity fields obtained
from DNS; one can consider, for example, random fields whose energy spectrum is
consistent with that of velocity fields obtained from DNS. The qualitative study of
Newtonian particles governed by   \eqref{e:stokes_intro} for given (random) velocity
fields is very similar to the theory of turbulent diffusion \cite{kramer}, which has been
primarily developed  in the case $\tau=0$. However, relatively little is known about the
properties of solutions in the inertial case. It is important, therefore, to consider
simplified models for the velocity field $v(x,t)$ which render the particle dynamics
amenable to rigorous mathematical analysis and careful numerical investigations.

A model for the motion of inertial particles in two dimensions was introduced in
\cite{inertial_2} and analyzed in a series of papers \cite{ HairPavl04,KupPavlSt04,
PavlSt03,PavSt05a,PavSt05b,PavlStBan06,inertial_1}. This model consists of motion in a
force field comprised of a contribution from Stokes' law together with molecular
bombardment; the velocity field is a Gaussian, Markovian, divergence--free random field.
This gives the equations
\begin{subequations}\label{e:motion}
\begin{equation}\label{e:stokes}
\tau \ddot{x}(t) = v(x(t),t) - \dot{x}(t) + \sigma \xi(t),
\end{equation}
\begin{equation} \label{e:skew_gradient}
v(x,t) = \nabla^{\bot} \phi(x,t),
\end{equation}
\begin{equation}\label{e:ou_inf}
\frac{\partial \phi}{\partial t} = A \phi + \sqrt{Q} \zeta(x,t).
\end{equation}
\end{subequations}
The parameter  $\tau$ is the Stokes number, which is a non--dimensional measure of the particle
inertia (essentially it is the particle relaxation time). The  molecular
diffusion coefficient is given  by  $\sigma \geq 0$
, $\xi(t)$ is white noise in $\R^2$, $\zeta(x,t)$ is space--time Gaussian white noise
and $A, \, Q$ are appropriate positive, self--adjoint operators. Gaussian velocity fields
of the form \eqref{e:ou_inf} have been considered by various authors in the past, in
particular in the context of passive advection. See for example \cite{CarCer99} and the
references therein. The usefulness of random velocity fields of the form \eqref{e:ou_inf}
in simulations is that, by choosing the operators $A$ and $Q$ appropriately, we can
generate random velocity fields with a given energy spectrum, thus creating caricatures
of realistic turbulent flows. Generalizations to arbitrary dimension $d$ are also possible

Various qualitative properties of the system \eqref{e:motion} have been studied, such as
existence and uniqueness of solutions and existence of a random attractor
\cite{inertial_1}. Furthermore, various limits of physical interest have been studied:
rapid decorrelation in time (Kraichnan) limits \cite{ KupPavlSt04,PavlSt03, PavSt05a}.

Diffusive
scaling limits (homogenization) were studied for time independent,
periodic in space velocity fields; thus the model is obtained
from \eqref{e:stokes} with $v(x,t)=v(x)$ only
\cite{HairPavl04, PavSt05b, PavlStBan06}. In these three
papers it was shown that the rescaled  process
\begin{equation}\label{e:resc}
x^\eps(t):= \eps x(t/\eps^2)
\end{equation}
converges in distribution  in the limit as $\eps \rightarrow 0$ to a Brownian motion with a
nonnegative definite effective diffusivity $\cK$. Various properties of the effective
diffusivity , in particular, the dependence of $\cK$ on the parameters of the
problem $\tau, \, \sigma$ were studied by means of formal asymptotics and extensive
numerical simulations. The purpose of this paper is to carry out a similar analysis for
the model problem \eqref{e:motion} where the velocity field is time-dependent.

When $\tau = 0$, i.e. the particle inertia is negligible, the equation of motion
\eqref{e:stokes} becomes
\begin{equation}\label{e:tracer}
\dot{x}(t) = v(x(t),t) + \sigma \xi(t).
\end{equation}
This equation has been studied extensively in the literature \cite{falko_verga,kramer}.
The homogenization problem for \eqref{e:tracer} with velocity fields of the form
 \eqref{e:skew_gradient},\eqref{e:ou_inf} was studied in \cite{carmona}. There it was shown that the rescaled
process \eqref{e:resc}, with $x(t)$ being the solution of \eqref{e:tracer} and $v(x,t)$
being a finite dimensional truncation of solutions to \eqref{e:skew_gradient},\eqref{e:ou_inf}, converges in
distribution to a Brownian motion with a positive definite covariance matrix, the {\it
effective diffusivity}.

In this paper we will show that a similar result holds for the inertial particles
problem. That is, we consider the diffusive rescaling \eqref{e:resc} for solutions to
\eqref{e:motion} with $v(x,t)$ being a Galerkin truncation of \eqref{e:ou_inf}. We show,
first with the aid of formal multiple scale expansions and then rigorously, that the
rescaled processes converges to a Brownian motion and we derive a formula for the
effective diffusion tensor. We study various properties of the effective diffusivity as
well as some scaling limits of physical interest. Furthermore, we analyze the dependence
of the effective diffusivity $\cK$ on the various parameters of the problem through numerical simulations. In
particular, we show that the effective diffusivity depends on the Stokes number $\tau$ in
a very complicated, highly nonlinear way; this leads to various interesting, physically
motivated, questions.

The generator of the Markov diffusion process corresponding to \eqref{e:motion} is not a
uniformly elliptic operator, as in the case of passive tracers, but a degenerate, {\it
hypoelliptic} operator. This renders the proof of the homogenization theorem for
\eqref{e:motion} quite involved, since rather sophisticated tools from the spectral
theory of hypoelliptic operators are required--see the Appendix.

The rest of the paper is organized as follows. In Section \ref{sec:model} we introduce
the exact model that we will analyze and present some of its properties. In Section
\ref{sec:mult_sc} we use the method of multiple scales to derive the homogenized
equation. In Section  \ref{sec:small_delta} we study simultaneously the problems of
homogenization and rapid decorrelation in time. In Section 5 we present the results of
numerical simulations. Section 6 is reserved for conclusions. The rigorous homogenization
theorem is stated and proved in the Appendix.
%
%
%
%
\section{The Model}
\label{sec:model} We will study the following model for the motion of an inertial particle
in $\R^d$ \cite{MaxRil83}
\begin{equation} \label{e:the_model}
\tau \ddot{x}(t)= v(x(t),t)-\dot{x}(t)+\sigma \xi(t),
\end{equation}
where $\tau, \, \sigma > 0$ and $\xi(t)$ is a standard white noise process on $\R^d$,
i.e. a mean zero generalized Gaussian process with
$$
\langle \xi_i(t) \xi_j(s) \rangle = \delta_{ij} \delta(t- s), \quad i, j = 1, \dots d.
$$
The velocity field $v(x,t)$ is of the form
\begin{equation}\label{e:vel}
v(x,t) = F(x) \mu(t),
\end{equation}
where for each fixed $x$,  $F(x): \R^n \rightarrow \R^d$ is an $n \times d$ matrix  smooth and  periodic
as a function of $x$, and $\mu (t)$ is a stationary generalized Ornstein--Uhlenbeck process on
$\R^n$:
\begin{equation}\label{e:ou_model}
\dot{\mu}(t) = - \frac{1}{\delta} A \mu(t) + \frac{1}{\delta} \sqrt{\Lambda} \zeta(t).
\end{equation}
Here $\zeta(t)$ is a standard Gaussian
white noise process on $\mathbb{R}^{n}$, which is
independent from $\xi(t)$, $\delta > 0$ and $A, \, \Lambda$ are positive definite $n
\times n$ matrices. The parameter $\delta$ controls the correlation time of the
Ornstein--Uhlenbeck process $\xi(t)$. We remark that one can construct a velocity field
\eqref{e:vel} through a finite dimensional truncation of \eqref{e:ou_inf}. Notice
however that we do not assume that the velocity field $v(x,t)$ is incompressible as such an
assumption is not needed for the analysis. We will, however, restrict ourselves to
incompressible velocity fields when studying the problem numerically in Section
\ref{sec:numerics}, as this case is physically interesting.

It is sometimes more convenient for the subsequent analysis to consider the rescaled OU process
$\eta(t) = \sqrt{\delta} \mu(t)$. Written in terms of $\eta(t)$, the equations that
govern the motion of inertial particles become
\begin{subequations}\label{e:motion_resc}
\begin{equation}\label{e:main_model}
\tau \ddot{x}(t)= \frac{F(x(t)) \eta(t)}{\sqrt{\delta}}-\dot{x}(t)+\sigma \xi(t),
\end{equation}
\begin{equation}\label{e:ou_model1}
\dot{\eta}(t) = - \frac{1}{\delta} A \eta(t) + \sqrt{\frac{\Lambda}{\delta}}  \zeta(t),
\end{equation}
\end{subequations}
The velocity field that appears in \eqref{e:main_model} is a mean zero stationary \footnote{For appropriately chosen initial conditions.} Gaussian  random field
with correlation time $\delta$.  It is possible to show \cite{PavlSt03}
that in the limit as $\delta \rightarrow 0$ (the rapid decorrelation in time limit) the
solution of \eqref{e:main_model} converges pathwise to the solution of
\begin{equation}\label{e:kraichnan}
\tau \ddot{x}(t) = F(x(t))A^{-1}\sqrt{\Lambda}\zeta(t) - \dot{x}(t) + \sigma \xi(t).
\end{equation}
The Kraichnan--like velocity field
$$v(x,t) = F(x)A^{-1}\sqrt{\Lambda}\zeta(t)$$
is mean zero, Gaussian, and delta--correlated in time. We will refer to  \eqref{e:motion_resc} as the \emph{colored velocity field model} and to
\eqref{e:kraichnan} as the \emph{white velocity field model}.

In this paper we will be mostly concerned with the diffusive limit of solutions to
\eqref{e:motion_resc} and \eqref{e:kraichnan} . That is, we will consider the rescaled process \eqref{e:resc} and study the limit as
$\eps \rightarrow 0$.  A natural question is whether the homogenization ($\eps
\rightarrow 0$) and rapid decorrelation in time ($\delta \rightarrow 0$) limits commute. We
answer this question in the affirmative through using formal asymptotics as well as
through numerical investigations.
%
%
%
%
\section{Multiple Scales Expansion for Effective Diffusivities}
\label{sec:mult_sc}
In this section we will derive the homogenized equation which describes the motion of
inertial particles at large length and  time scales for both the colored and white noise velocity fields.
The derivation of the homogenized equation is based on multiscale/homogenization techniques
\cite{lions}. We refer to \cite{PavlSt06b}  for a recent pedagogical introduction to such methods.
\subsection{Homogenization for the colored velocity field}
We start by rescaling the equations of motion \eqref{e:motion_resc} according to $t
\mapsto t/\epsilon^{2}$, $x \mapsto x/\epsilon$. Using the fact that for any white noise
process we have that $\xi(ct)=\frac{1}{\sqrt{c}}\xi(t)$ in law we obtain:
\begin{subequations}
\label{e:color_mult_resc}
\begin{eqnarray}
\tau \epsilon^{2}\ddot{x}^{\eps} &=& \frac{1}{\epsilon}F \left(\frac{x^{\eps}}{\epsilon}
\right)\frac{\eta^{\eps}}{\sqrt{\delta}} -\dot{x}^{\eps}+\sigma \xi, \\
\dot{\eta^{\eps}} & = & -\frac{1}{\epsilon^{2}}\frac{A}{\delta}\eta^{\eps} +
\frac{1}{\epsilon}\sqrt{\frac{\Lambda}{\delta}}\zeta.
\end{eqnarray}
\end{subequations}
We now introduce two new variables $y^{\eps}=\sqrt{\tau} \epsilon \dot{x}^{\eps}$ and $z^{\eps}=x^{\eps}/\epsilon$
and write the above equations as a first order system:
\begin{subequations} \label{e:sde_resc}
\begin{eqnarray}
d x^{\eps}(t) &=& \frac{1}{\sqrt{\tau}\epsilon} y^{\eps}(t) \, dt, \\
d y^{\eps}(t) &=& \frac{1}{\epsilon^{2}\sqrt{\tau}} \frac{F(z^{\eps}(t)) \eta^{\eps}(t)}{\sqrt{\delta}} \, dt
-\frac{1}{\tau \epsilon^{2}}y^{\eps}(t) \, dt+
\frac{\sigma}{\sqrt{\tau}\epsilon} \,  dB(t), \\
d z^{\eps}(t) &=& \frac{1}{\sqrt{\tau} \epsilon^{2}}y^{\eps}(t) \, dt, \\
d \eta^{\eps}(t)&=&-\frac{1}{\epsilon^{2}}\frac{A}{\delta}\eta^{\eps}(t) \, dt +\frac{1}{\epsilon}
\sqrt{\frac{\Lambda}{\delta}} \, dW(t),
\end{eqnarray}
\end{subequations}
with the understanding that $z^{\eps} \in \mathbb{T}^{d}, \, x^{\eps},y^{\eps} \in \mathbb{R}^{d}, \, \eta^{\eps} \in
\mathbb{R}^{n}$ and with $\xi(t) = \dot{B}(t), \, \zeta(t) = \dot{W}(t)$, $B(t)$ and
$W(t)$ being standard Brownian motions on $\R^d$ and $\R^n$, respectively. The SDEs
\eqref{e:sde_resc} clearly exhibit the two time scales
$\mathcal{O}(\epsilon)$ (for $x^{\eps}$) and $\mathcal{O}(\epsilon^{2})$ (for
$(y^{\eps},z^{\eps},\eta^{\eps})$). Our purpose is to homogenize over
the fast variables $(y^{\eps},z^{\eps},\eta^{\eps})$ to obtain a
closed equation which governs
the evolution of $x(t)$, and is valid for $\eps \ll 1$.
%
%
%
%
%
%
\subsubsection {Multiscale expansion}
Let $X^{x,y,z,\eta}_t :=\{x^{\eps}(t), y^{\eps}(t), z^{\eps}(t), \eta^{\eps}(t) \}$ denote the solution of
\eqref{e:sde_resc} starting at $ \{x,\,y,\,z,\,\eta \}$ and let $f:\R^d \times \R^d \times \T^d
\times \R^n \mapsto \R$ be a smooth function. Then the observable $u^\eps(x,y,z,\eta, t)
= \IE f(X^{x,y,z,\eta}_t)$ satisfies the backward Kolmogorov equation associated with the
rescaled process:
\begin{eqnarray} \label{e:kolm_resc}
\frac{\partial u^{\epsilon}}{\partial t} &=& \frac{1}{\eps} \left( \frac{1}{\sqrt{\tau}}y
\cdot \nabla_{x} \right) u^{\epsilon} +\frac{1}{\epsilon^{2}}\left[
\frac{1}{\sqrt{\tau}}y \cdot \nabla_{z} \nonumber
+\frac{1}{\sqrt{\tau}}\frac{F(z) \eta}{\sqrt{\delta}} \cdot \nabla_{y} \right. \\
&& \left.+\left(-\frac{A}{\delta}\eta \cdot
\nabla_{\eta}+\frac{\Lambda}{2\delta} :
 \nabla_{\eta}\nabla_{\eta} \right) +\frac{1}{\tau}\left(-y \cdot \nabla_{y}+\frac{\sigma^{2}}{2} \Delta_{y}\right)
 \right]u^{\epsilon} \nonumber \\
&=:& \left( \frac{1}{\epsilon^{2}}\mathcal{L}_{0}+\frac{1}{\epsilon} \mathcal{L}_{1}
\right) u^{\epsilon}, \ \text{with} \ \ u^{\eps}|_{t=0}=f.
\end{eqnarray}
Here:
\begin{subequations}\label{e:oper_defn}
\begin{eqnarray}
\mathcal{L}_{0} &:=& \frac{1}{\sqrt{\tau}} \left( y \cdot \nabla_{z}+\frac{F(z)
\eta}{\sqrt{\delta}}\cdot \nabla_{y} \right)+
\frac{1}{\tau}\mathcal{L}_{OU,y}+\frac{1}{\delta}\mathcal{L}_{OU, \eta}, \label{e:oper_defn_lo} \\
\mathcal{L}_{1} &:=&\frac{1}{\sqrt{\tau}} y \cdot \nabla_{x}, \\
\mathcal{L}_{OU, \eta}
& := & -A\eta \cdot \nabla_{\eta}+\frac{\Lambda}{2} :
 \nabla_{\eta}\nabla_{\eta},   \\
\mathcal{L}_{OU,y} & := & -y \cdot \nabla_{y}+\frac{\sigma^{2}}{2} \Delta_{y}.
\end{eqnarray}
\end{subequations}
We use the subscript $OU$ on the last two operators to emphasize that they
are the generators of Ornstein-Uhlenbeck processes in $\eta$ and $y$
respectively.
The operator $\mathcal{L}_{0}$ is the generator of the Markov process $\{z(t), y(t),
\eta(t) \} \in \T^d \times \R^d \times \R^n$:
\begin{subequations} \label{e:fast_proc}
\begin{eqnarray}
d z(t)& = &\frac{1}{\sqrt{\tau}}y(t) \, dt, \\
d y(t)& = &\frac{1}{\sqrt{\tau}} \frac{ F(z(t)) \eta(t)}{\sqrt{\delta}} \, dt -
\frac{1}{\tau}y(t) \, dt +
\frac{\sigma}{\sqrt{\tau}} \, dB(t), \\
d \eta(t)& = &-\frac{A}{\delta}\eta(t) \, dt+ \sqrt{\frac{\Lambda}{\delta}} \,
dW(t).
\end{eqnarray}
\end{subequations}
In the Appendix we prove that the Markov process $\{z(t), y(t), \eta(t) \}$ given by \eqref{e:fast_proc} is
geometrically ergodic. Hence, there exists a unique invariant measure with smooth density
$\rho(z,y, \eta)$ which is the solution of the stationary Fokker--Planck equation
\begin{equation} \label{e:ff_stat}
\mathcal{L}_{0}^{\ast}\rho =  -\frac{1}{\sqrt{\tau}} \left( y \cdot \nabla_{z}\rho
+F(z)\eta\cdot \nabla_{y}\rho \right)+\frac{1}{\tau} \cL_{OU,y}^* \rho + \frac{1}{\delta}\cL_{OU,\eta}^*
\rho
  = 0.
\end{equation}
Here
$$
\cL_{OU,y}^* \cdot := \nabla_y \cdot (y\cdot) + \frac{\sigma^2}{2} \Delta_y \cdot \quad
\mbox{and} \quad \cL_{OU,\eta}^* \cdot :=  \nabla_{\eta}\cdot(A
\eta \cdot) + \frac{\Lambda}{2}:\nabla_{\eta} \nabla_{\eta} \cdot
$$
are the formal $L^2$--adjoints of $\cL_{OU,y}$ and $\cL_{OU,\eta}$, respectively. The
invariant density $\rho(z,y,\eta)$ is a periodic function of $z$ and decays rapidly as
$\| z \|, \, \| y \| \rightarrow \infty$.

In the Appendix we also prove that the operator $\LL^*_0$ (equipped with boundary
conditions described above) has compact resolvent in the appropriate function space.
Consequently, Fredholm theory applies: the null space of the generator $\mathcal{L}_{0}$
is one dimensional and consists of constants in $z,y,\eta$. (See the
Appendix). Moreover the equation
$\mathcal{L}_{0}f=g$, has a unique (up to constants) solution if and only if
\begin{equation*}
\left\langle g \right\rangle_{\rho} :=\int_{\mathbb{X}} g(y,z,\eta) \rho(y,z,\eta)\, dX=0
\end{equation*}
where $\mathbb{X}:=\mathbb{T}^{d}\times \IR^{d}\times \IR^{n}$ and $dX := dydzd\eta$. We
assume that that the average of the velocity with respect to the invariant density $\rho$
vanishes:
\begin{equation} \label{e:centering}
\left\langle F(z)\eta \right\rangle_{\rho}=0.
\end{equation}
This is natural because it removes any effective  drift making a purely diffusive scaling natural.
The identity $\int_{\mathbb{X}}y\big(\mathcal{L}_{0}^{\ast}\rho(y,z,\eta)\big) \, dX=0$ implies
that
\begin{equation*}
\int_{\mathbb{X}}\big(\mathcal{L}_{0}y\big)\rho(y,z,\eta) \,dX=0.
\end{equation*}
Consequently:
\begin{equation*}
\int_{\mathbb{X}}\left( \frac{1}{\sqrt{\tau}}\frac{F(z)\eta}
{\sqrt{\delta}}-\frac{1}{\tau}y \right) \rho(y,z,\eta) \, dX=0
\end{equation*}
Thus, the centering condition \eqref{e:centering} is equivalent to:
\begin{equation}\label{e:centering1}
\left\langle y \right\rangle_{\rho}=0.
\end{equation}

Let us now proceed with the derivation of the homogenized equation. We look for a solution of
\eqref{e:kolm_resc} in the form of a power series in $\eps$:
\begin{equation} \label{e:ansatz}
u^{\epsilon}=u_{0} (x,y,z,\eta,t)+\epsilon u_{1} (x,y,z,\eta,t)+\epsilon^{2} u_{2}(x,y,z,\eta,t)
+ \ldots
\end{equation}
with $u_{i}=u_{i}(x,y,z,\eta,t), \, i=1,2,\ldots$ being $1$--periodic in $z$.  We
substitute \eqref{e:ansatz} into (\ref{e:kolm_resc}) and obtain the following sequence of
equations:
\begin{subequations} \label{e:e0e1e2}
\begin{eqnarray}
-\mathcal{L}_{0}u_{0}& = &0,  \label{e:e0} \\
-\mathcal{L}_{0}u_{1}& = & \mathcal{L}_{1}u_{0},  \label{e:e1} \\
-\mathcal{L}_{0}u_{2}& = &  \mathcal{L}_{1}u_{1} - \frac{\partial u_{0}}{\partial t},
\label{e:e2}
\end{eqnarray}
\end{subequations}
From \eqref{e:e0}, we deduce that the first term in the expansion is independent of the
fast variables $z,\, y, \, \eta$, i.e. $u_{0}=u(x,t)$. Now equation \eqref{e:e1} becomes:
\begin{equation*}
-\mathcal{L}_{0}u_{1}=\frac{1}{\sqrt{\tau}}y\cdot \nabla_{x}u.
\end{equation*}
We can solve this by using separation of variables\footnote{In
principle we should also add a term which is in the null space of $\LL_0$. It is easy to
show, however, that this term does not affect the homogenized equation and for simplicity
we set it equal to $0$.}:
\begin{equation*}
u_{1}=\Phi(y,z,\eta) \cdot \nabla_{x}u,
\end{equation*}
with
\begin{equation} \label{e:cell_color}
-\mathcal{L}_{0}\Phi(y,z,\eta)=\frac{1}{\sqrt{\tau}}y.
\end{equation}
This is the \emph{cell problem} which is posed on $ \mathbb{X}$. Assumption \eqref{e:centering1} implies that the right hand side of the the above
equation is centered with respect to the invariant measure of the fast process. Hence the
equation is well posed (see the Appendix). The boundary conditions
for this PDE are that $\Phi(y,z,\eta)$ is periodic in $z$ and that it belongs to
$L^2(\mathbb{X}, \rho(z,y,\eta) \, d z d y d \eta)$, which implies sufficiently fast decay at infinity.

We  now proceed with \eqref{e:e2}. We apply the solvability condition to obtain:
\begin{eqnarray}
\frac{\partial u_{0}}{\partial t} & = & \left\langle \mathcal{L}_{1}u_{1} \right\rangle_{\rho}
\nonumber \\
& = & \frac{1}{\sqrt{\tau}}\left\langle y\otimes\Phi \right\rangle_{\rho}:
\nabla_{x}\nabla_{x}u.
\end{eqnarray}
Thus, the backward Kolmogorov equation which governs the dynamics on large scales is
\begin{equation} \label{e:kolmog_homog}
\frac{\partial u}{\partial t}= \mathcal{K}: \nabla_{x}\nabla_{x}u,
\end{equation}
where the effective diffusivity is given by
\begin{equation}\label{e:eff_diff_form}
\mathcal{K}=\frac{1}{\sqrt{\tau}}\left\langle y \otimes \Phi \right\rangle_{\rho},
\end{equation}
and where $\otimes$ denotes the tensor (or outer) product.
Notice that only the symmetric part of the effective diffusivity  $\mathcal{K}$ is relevant in  the
homogenized equation \eqref{e:kolmog_homog}. However, the effective diffusivity $\mathcal{K}$ itself   is
non--symmetric in general. We define
$$
sym(\cK)=\frac{1}{2}(\cK +\cK^{T})
$$

\subsubsection{The homogenization result}

Equation \eqref{e:kolmog_homog} is the Backward Kolmogorov equation
corresponding to a pure Brownian motion. We have the following result:
\result
For $\eps \ll 1$ and $t=\mathcal{O}(1)$
the function $x^{\eps}(t)=
\eps x(t/\eps^2)$, where $x(t)$ solves
\eqref{e:motion_resc},
is approximated by $X(t)$ solving
\begin{equation}
\label{e:sde_homog}
\dot{X}=\sqrt{2 sym(\cK)} \dot{\beta}.
\end{equation}
with $X(0)=x(0),$ and where $\beta$ is a standard Brownian motion on $\bbR^d.$
\normalfont
\vspace{1cm}

A theorem, justifying the formal approximation leading to this result,
is proved in the Appendix, using the martingale central limit theorem.

\subsubsection{Properties of the effective diffusivity }
In this subsection we show that the effective diffusivity is non negative. This implies that the homogenization equation is well posed.  We can show this  by using the Dirichlet form (Theorem 6.12 in \cite{PavlSt06b}) which shows that, for
every sufficiently smooth $f(z,y,\eta)$,
\begin{equation}\label{e:dirichlet}
\int_{\mathbb{X}}f(- \mathcal{L}_{0}f)\rho \, dX =\frac{\sigma^{2}}{2\tau}\left\langle
|\nabla_{y}f|^{2} \right\rangle_{\rho} +\frac{1}{2\delta}\left\langle
(\nabla_{\eta}f)^{T} \Lambda(\nabla_{\eta}f) \right\rangle_{\rho}.
\end{equation}

Now let $\Phi$ be the solution of the Poisson equation \eqref{e:cell_color} and define
$\phi^{\alpha} :=\alpha \cdot \Phi$, where $\alpha \in \R^d$ is an arbitrary unit vector.
The scalar field $\phi^{\alpha}$ satisfies the Poisson equation
\begin{equation*}
-\mathcal{L}_{0}\phi^\alpha=\frac{1}{\sqrt{\tau}}\alpha \cdot y.
\end{equation*}
We combine \eqref{e:eff_diff_form} with  \eqref{e:dirichlet}, to calculate
\begin{eqnarray}
\alpha. \mathcal{K}\alpha &=& \frac{1}{\sqrt{\tau}}\int_{\mathbb{X}}(\alpha \cdot y)
(\alpha \cdot \Phi)\rho \, dX \nonumber \\
&=&\int_{\mathbb{X}}\phi^{\alpha} (- \mathcal{L}_{0}\phi^{\alpha}) \rho \, dX \nonumber
\\ &=&\frac{\sigma^{2}}{2\tau}\left\langle |\nabla_{y}\phi^\alpha|^{2}
\right\rangle_{\rho}+\frac{1}{2\delta} \left\langle
(\nabla_{\eta}\phi^\alpha)^{T}\Lambda(\nabla_{\eta}\phi^\alpha) \right\rangle_{\rho}
\label{e:eff_diff_2} \\
\nonumber & \geq & 0,
\end{eqnarray}
since $\Lambda$ is a positive definite matrix. Thus the following results holds:
\result
The effective diffusivity matrix $\mathcal{K}$ given by  \eqref{e:eff_diff_form} is
positive semi-definite and thus the limiting backward Kolmogorov equation \eqref{e:kolmog_homog} is well posed.
\rem \normalfont It is not entirely straightforward to check whether the centering
condition \eqref{e:centering} or, equivalently, \eqref{e:centering1} is satisfied or not,
as we don't have a formula for the invariant measure of the fast process--we only know
that it exists. It is possible, however, to identify some general classes of flows
$v(x,t)$ which satisfy \eqref{e:centering} by using symmetry arguments. Consider for
example  the case of a parity invariant flow, i.e a flow satisfying the condition
\begin{equation} \label{e:parity}
F(-z)=-F(z).
\end{equation}
It follows from \eqref{e:parity} and \eqref{e:ff_stat} that the invariant density
satisfies
\begin{equation}\label{e:rho_symm}
\rho(y,z,\eta)=\rho(-y,-z,\eta).
\end{equation}
It is easy to see now that \eqref{e:rho_symm} implies that \eqref{e:centering} is
satisfied. Hence, the centering condition is satisfied for velocity fields $v(x,t)$ that
are odd functions of $x$.
\normalfont
\rem \label{rem:drift} \normalfont Even if the centering condition is not satisfied, the
large-scale, long-time dynamics of the inertial particle is still governed by an
effective Brownian motion, provided that we study the problem in a frame co-moving with
the mean flow. Indeed, if we denote by $V$ the mean flow, i.e.
$$
V = \langle F(z) \eta \rangle_{\rho},
$$
then the rescaled processed $x^{\eps} = \eps \left( x(t/\eps^2) - V t/\eps^2 \right)$
converges in distribution to a Brownian motion with covariance matrix (the effective
diffusivity) given by
\begin{equation}\label{e:deff_drift}
\cK = \frac{1}{\sqrt \tau}
\left\langle \left(y - \sqrt{\frac{\tau}{\delta}}V \right) \otimes \Psi
\right\rangle_{\rho}
\end{equation}
with
\begin{equation}\label{e:cell_drift}
-\LL_0 \Psi = \frac{1}{\sqrt \tau}\Bigl(y - \sqrt{\frac{\tau}{\delta}} V\Bigr),
\quad \LL_0^* \rho = 0.
\end{equation}

\subsection{Homogenization: the white noise velocity field}
We can use the same multiscale techniques to study the diffusive scaling of
\eqref{e:main_model} for the white velocity field, equation \eqref{e:kraichnan}:
\begin{equation*}
\tau \ddot{x}(t)=F(x(t))A^{-1}\sqrt{\Lambda}\zeta(t)-\dot{x}+\sigma \xi(t).
\end{equation*}
After  a similar calculation with the colored noise problem  we find that the backward Kolmogorov equation which governs the dynamics on large scale is
\begin{equation} \label{e:kolmog_homog1}
\frac{\partial u}{\partial t}=\widehat{\cK}: \nabla_{x}\nabla_{x}u,
\end{equation}
where the effective diffusivity is given by
\begin{equation}\label{e:eff_diff:white}
 \widehat{\cK} = \frac{1}{\sqrt{\tau}} \langle y \otimes \widehat{\Phi}
 \rangle_{\widehat{\rho}}.
\end{equation}
Here
\begin{equation} \label{e:cell_white}
-\widehat{\LL}_{0}\widehat{\Phi}(y,z)=\frac{1}{\sqrt{\tau}}y, \quad \widehat{\cL}_0^*
\widehat{\rho} = 0,
\end{equation}
with
\begin{equation}\label{e:l_white}
\widehat{\mathcal{L}}_{0} := \frac{1}{\sqrt{\tau}}y \cdot \nabla_{z}+\frac{1}{2\tau}F(z)
A^{-1}\Lambda A^{-1}F^T(z): \nabla_{y} \nabla_{y}+\frac{1}{\tau}\mathcal{L}_{OU,y}.
\end{equation}
The operator $\cL_{OU,y}$ is defined in \eqref{e:oper_defn}. We use the notation $\langle
\rangle_{\widehat{\rho}}$ to denote averaging over $\T^d \times \R^d$ with respect to the
invariant distribution $\widehat{\rho}$.
\subsubsection{The homogenization result}
Equation \eqref{e:kolmog_homog1} is the backward Kolmogorov
equation corresponding to a pure Brownian motion.
Hence we have the following result:
\result
For $\eps <<1$ and $t=\mathcal{O}(1)$
the function $x^{\eps}(t)=
\eps x(t/\eps^2)$, where $x(t)$ solves
\eqref{e:kraichnan}, is approximated by $X(t)$ solving
\begin{equation}
\label{e:thisone}
\dot{X}=\sqrt{2 sym(\widehat{\cK})} \dot{\beta},
\end{equation}
with $X(0)=x(0),$ and where $\beta$ is a standard Brownian motion on $\bbR^d.$
\normalfont
\vspace{1cm}

This result can be justified rigorously by means of the martingale
central limit theorem, as is done for the coloured noise case in
the Appendix.
\subsubsection{Properties of the effective diffusivity}
As in the case of the colored velocity field, the covariance matrix of the limiting
Brownian motion is nonnegative define. Indeed, let $\alpha \in \R^d$ be an arbitrary unit
vector, define $ \widehat{\phi}^{\alpha}:= \alpha \cdot \widehat{\Phi}$ and use the Dirichlet form (Theorem 6.12 in \cite{PavlSt06b})
\begin{eqnarray*}
\alpha.\widehat{\cK} \alpha &=&\frac{\sigma^{2}}{2\tau}\left \langle |\nabla_{y}
\widehat{\phi}^\alpha|^{2} \right \rangle_{\widehat{\rho}} + \frac{1}{2\tau}\left \langle
(\nabla_{y} \widehat{\phi}^\alpha) ^{T}F(z)A^{-1}\Lambda A^{-1}F(z)^{T} (\nabla_{y}
\widehat{\phi}^\alpha)\right \rangle_{\widehat{\rho}}
\\ & \geq & 0.
\end{eqnarray*}
Thus we have the following result:

\result The effective diffusivity matrix $\widehat{\cK}$
given by \eqref{e:eff_diff:white} is positive semi-definite and thus the associated
backward Kolmogorov equation is well posed.
\rem \normalfont An important observation is that the centering condition
for the white noise problem is always satisfied. Indeed, let $\widehat{\bX} := \T^d \times \R^d$, $d \widehat{X}:= dzdy$ and use the
identity
$$\int_{\widehat{ \mathbb{X}}}y\widehat{\LL}_{0}^{\ast}\widehat{\rho}(y,z) d\widehat{X}=0$$
together with integrations by parts to deduce that
\begin{equation*}
\int_{\widehat{\mathbb{X}}}y \widehat{\rho}(y,z) d\widehat{X}=0.
\end{equation*}
This suffices for solvability of \eqref{e:cell_white}.
Hence, the long--time, large--scale behavior of solutions to \eqref{e:kraichnan} is
always diffusive. This is to be contrasted with the case of the colored velocity field,
where  an additional condition, equation \eqref{e:centering}, has to be imposed to
ensure diffusive large scale dynamics.
\normalfont
%
\section{White Noise Limit for the Effective Diffusivity}
\label{sec:small_delta}
Consider the rescaled equation~\eqref{e:color_mult_resc}, and denote its solution by
$x^{\delta, \eps}(t)$. It is clear that if we first take the limit as $\delta \rightarrow
0$ and then the limit $\eps \rightarrow 0$, then $x^{\delta, \eps}(t)$ converges to a
Brownian motion with covariance matrix $\widehat{\cK}$ given by
eqn.~\eqref{e:eff_diff:white}, without having to impose any centering condition. A
natural question arises as to what happens if we interchange the order with which we take
the limits $\delta,\epsilon \to 0$.
In this section we show that the two limits commute under the additional
assumption that the centering condition \eqref{e:centering} is satisfied.
 In particular we have the following result:
 \result
\label{r:inter}
 Let $A, \Lambda$ be positive definite matrices that commute and assume that the centering
 condition \eqref{e:centering} is satisfied. Then for
 $\delta \ll 1$ the effective diffusivity $\cK$ from the colored noise model
   given by  \eqref{e:eff_diff_form}  admits the asymptotic
expansion
\begin{equation}\label{e:eff_diff_delta_exp}
\cK = \widehat{\cK} + \mathcal{O}(\sqrt{\delta}),
\end{equation}
where $\widehat{\cK}$  is given by \eqref{e:eff_diff:white}.
\normalfont \newline

The derivation of \eqref{e:eff_diff_delta_exp} is based on singular perturbation analysis
of the cell problem \eqref{e:cell_color} and of the stationary Fokker-Planck equation
\eqref{e:ff_stat}; see \cite{pavl05, PavSt05b,HorsLef84}. We start by writing the
operator $\mathcal{L}_{0}$ defined in \eqref{e:oper_defn} in the form,
\begin{equation} \label{e:modified_l}
\mathcal{L}_{0}=\frac{1}{\delta}\mathcal{A}_{0}+\frac{1}{\sqrt{\delta}}\mathcal{A}_{1}+
\mathcal{A}_{2},
\end{equation}
with
\begin{equation*}
\mathcal{A}_{0}=\mathcal{L}_{OU, \eta},
\quad \mathcal{A}_{1}=\frac{1}{\sqrt{\tau}}F(z)\eta \cdot \nabla_{y}, \quad
\mathcal{A}_{2}= \frac{1}{\sqrt{\tau}}y\cdot \nabla_z
+\frac{1}{\tau}\mathcal{L}_{OU,y}.
\end{equation*}
Of course, $\cL_0^{\ast}$ is also of the form \eqref{e:modified_l} with $\cA_j$ replaced
with $\cA_j^*, \, j=0,1,2$.

Note that $\mathcal{A}_{0}$ is the generator of a $d$-dimensional OU process. Hence,
it has a one--dimensional null space which consists of constants in $\eta$. Furthermore, the
process generated by $\cA_0$ is geometrically ergodic and its invariant measure is
Gaussian. Since $A$ and $\Lambda$ commute, the
density of the unique Gaussian invariant measure (i.e., the solution of the equation
$\cA_0^* \rho^{\eta} = 0$) is
$$
\rho^{\eta} = \frac{1}{Z} \exp \bigl(- \eta^T \Lambda^{-1} A \eta \bigr),
$$
where $Z$ is the normalization constant.

Let $\Phi$ be the solution of \eqref{e:cell_color}. As before, we define $\phi^{\alpha} =
\Phi \cdot \alpha$ for an arbitrary unit vector $\alpha \in \R^d$. We have that
\begin{equation}\label{e:cell_psi}
- \cL_0 \phi^\alpha = \frac{1}{\sqrt{\tau}} y \cdot \alpha,
\end{equation}
that
\begin{equation}\label{e:fp_rhobar}
\cL_0^{\star} \rho = 0
\end{equation}
and that
\begin{equation}\label{e:eff_diff_psi}
\alpha \cdot \cK \alpha = \frac{1}{\sqrt{\tau}} \langle \alpha \cdot y \phi^\alpha
\rangle_{\rho}
\end{equation}
Now we need to calculate the small $\delta$ asymptotics of $\phi^\alpha$ and $\rho$.
%
\subsection{Expansion for $\phi^{\alpha}$}
We look for a solution of \eqref{e:cell_psi} in the form of a power series in
$\sqrt{\delta}$.
$$
\phi^{\alpha}=\phi^{\alpha}_{0}+\sqrt{\delta} \phi^{\alpha}_{1}+\delta \phi^{\alpha}_{2}
+ \dots
$$
We substitute the above into  \eqref{e:cell_psi} to obtain the following sequence of
equations
\begin{subequations} \label{e:phie0e1e2}
\begin{eqnarray}
-\mathcal{A}_{0}\phi^{\alpha}_{0}& = &0, \label{e:phie0} \\
-\mathcal{A}_{0}\phi^{\alpha}_{1}& = & \mathcal{A}_{1}\phi^{\alpha}_{0}, \label{e:phie1}
\\
-\mathcal{A}_{0}\phi^{\alpha}_{2}& = &
\mathcal{A}_{1}\phi^{\alpha}_{1}+\mathcal{A}_{2}\phi^{\alpha}_{0}+
\frac{1}{\sqrt{\tau}}y \cdot \alpha. \label{e:phie2}
\end{eqnarray}
\end{subequations}
From Equation \eqref{e:phie0} we get that $\phi^{\alpha}_{0}=\phi^{\alpha}_{0}(z,y)$. In
order for equation~\eqref{e:phie1} to be well posed it is necessary that the right hand
side of the equation is orthogonal to the null space of $\cA_0^*$, i.e. that
$$
\left \langle \frac{1}{\sqrt{\tau}}F(z)\eta \cdot \nabla_{y}\phi_{0}^{\alpha}(z,y) \right
\rangle_{\rho^{\eta}} = 0,
$$
which is satisfied, since the term to be averaged is linear in $\eta$ and $\rho^{\eta}$ is a mean zero Gaussian density. The solution of
Equation \eqref{e:phie1} is
$$
\phi^\alpha_{1}=\frac{1}{\sqrt{\tau}}F(z)A^{-1}\eta \cdot
\nabla_{y}\phi^{\alpha}_{0}+\widehat{\Psi}_{1}(z,y).
$$
The solvability condition for \eqref{e:phie2} gives
\begin{eqnarray*}
\left \langle \frac{1}{\tau}\eta^{T} \Big( F^T(z)\nabla_{y}\nabla_{y}
 \phi^{\alpha}_{0}F(z)A^{-1} \Big)\eta   \right \rangle_{\rho^{\eta}}+
 \frac{1}{\sqrt{\tau}} y \cdot \nabla_z \phi^{\alpha}_{0}+\frac{1}{\tau}
 \mathcal{L}_{OU,y}\phi^{\alpha}_{0} +\frac{1}{\sqrt{\tau}}y\cdot \alpha = 0.
\end{eqnarray*}
We use the fact that
$$\langle \eta^{T}B\eta \rangle_{\rho^{\eta}}=\frac12 B: A^{-1}\Lambda
\quad \forall \, B \in \R^{d \times d},$$ to obtain
\begin{equation} \label{e:cell_psi0}
- \widehat{\cL}_0 \phi^{\alpha}_{0}= \frac{1}{\sqrt{\tau}} y \cdot \alpha.
\end{equation}
This is precisely the cell problem for the white noise velocity field,
equation \eqref{e:cell_white} projected along the direction $\alpha \in \R^d$. Hence, the
small $\delta$ expansion of the solution to \eqref{e:cell_psi} is
\begin{equation}\label{e:psi_exp}
\phi^{\alpha}(z,y, \eta) = \phi^{\alpha}_0(z,y) + \sqrt{\delta}
\left(\frac{1}{\sqrt{\tau}}F(z)A^{-1}\eta \cdot
\nabla_{y}\phi^{\alpha}_{0}+\widehat{\Psi}_{1}(z,y) \right) + \cO(\delta),
\end{equation}
where $\phi^{\alpha}_0(z,y)$ is the solution to \eqref{e:cell_psi0}.

%
%
\subsection{Expansion for $\rho$}
We look for a solution of \eqref{e:fp_rhobar} in the form of a power series in
$\sqrt{\delta}$
$$
\rho=\rho_{0}+\sqrt{\delta} \rho_{1}+\delta\rho_{2} + \dots
$$
We substitute this expansion into \eqref{e:fp_rhobar} and equate equal powers of $\delta$
to obtain the following sequence of equations.
\begin{subequations} \label{e:rhoe0e1e2}
\begin{eqnarray}
-\mathcal{A}_{0}^{*}\rho_{0} & =&0,  \label{e:rhoe0}\\
-\mathcal{A}_{0}^{*}\rho_{1}& = & \mathcal{A}_{1}^{*}\rho_{0}, \label{e:rhoe1} \\
-\mathcal{A}_{0}^{*}\rho_{2}& = & \mathcal{A}_{1}^{*}\rho_{1}+\mathcal{A}_{2}^{*}\rho_{0} .\label{e:rhoe2}
\end{eqnarray}
\end{subequations}
From the first equation we deduce that (abusing notation)
$\rho_{0}(z,y,\eta)=\rho_0(z,y)\rho^{\eta}$. The solvability
condition for \eqref{e:rhoe1} is satisfied since
$$
-\int_{\mathbb{R}^{m}} \mathcal{A}_{1}^{*}\rho_{0} \, d \eta =  \int_{\mathbb{R}^{m}}
\frac{1}{\sqrt{\tau}}\Bigl(F(z)\eta \cdot \nabla_{y}\rho_0(z,y)\Bigr)
 \rho^{\eta} \,  d\eta = 0.
$$
The solution of \eqref{e:rhoe1} is
$$
\rho_{1}(z, y, \eta)=\rho^{\eta} \left(\frac{1}{\sqrt{\tau}}F(z)A^{-1}\eta
\cdot\nabla_{y}\rho_0(z,y) +\overline{\rho}_1(z,y) \right).
$$
The solvability condition for \eqref{e:rhoe2} is
\begin{equation}\label{e:solv_rho}
\int_{\R^m} \left( \mathcal{A}_{1}^{*}\rho_{1}+\mathcal{A}_{2}^{*}\rho_{0} \right) \, d
\eta = 0.
\end{equation}
We use the expressions for $\rho_0$ and $\rho_1$ to deduce that
$$
\int_{\R^m} \cA_2^* \rho_0(z,y,\eta)  \, d \eta=
-\frac{1}{\sqrt{\tau}}y\cdot \nabla_{z}\rho_{0}(z,y)+
\frac{1}{\tau}\mathcal{L}_{OU,y}^{\ast}\rho_0(z,y).
$$
and
$$
\int_{\R^m} \cA_1^* \rho_1  \, d \eta=\frac{1}{2\tau}F(z)A^{-1}\Lambda A^{-1}F^{T}(z):\nabla_{y}\nabla_{y}\rho_{0}.
$$
We substitute the above expressions in \eqref{e:solv_rho} to conclude that $\rho_0(z,y)$
satisfies
\begin{equation}\label{e:rho0}
\widehat{\cL}_0^{\star} \rho_0 = 0.
\end{equation}
Consequently, $\rho_0(z,y) = \widehat{\rho}(z,y)$, the solution of  the second equation in \eqref{e:cell_white}. Thus, the small $\delta$ expansion of $\rho$, the solution of
\eqref{e:fp_rhobar} is
\begin{equation}\label{e:rho_exp}
\rho = \widehat{\rho}(z,y)\rho^{\eta} + \sqrt{\delta} \rho^{\eta} \left(\frac{1}{\sqrt{\tau}}F(x)A^{-1}\eta
\cdot\nabla_{y}\rho_0(z,y) + \overline{\rho}_1(z,y) \right) + \cO(\delta).
\end{equation}
%
%
\subsection{Proof of Result \ref{r:inter}}
We have that
\begin{equation} \label{e:ef_dif_alpha}
\alpha \cdot \cK \alpha = \frac{1}{\sqrt{\tau }} \langle \alpha \cdot y
\phi^{\alpha} \rangle_{\rho}.
\end{equation}
In the previous subsections we expanded $\phi^{\alpha}$ and $\rho$ as
\begin{eqnarray*}
\phi^{\alpha} &=& \phi^{\alpha}_{0}+\sqrt{\delta} \phi^{\alpha}_{1}+\delta \phi^{\alpha}_{2}+\cdots \\
\rho &=&\rho_{0}+\sqrt{\delta} \rho_{1}+\delta\rho_{2} + \dots
\end{eqnarray*}
and showed that
\begin{eqnarray*}
\phi^{\alpha}_{0}(x,y,\eta) &=& \phi_{0}^{\alpha}(x,y), \\
\rho_{0}(x,y,\eta)  &=& \rho^{\eta}\widehat{\rho}(x,y),
\end{eqnarray*}
where $\phi_0^{\alpha}={\widehat \Phi}\cdot \alpha$ and
${\widehat \Phi}, {\widehat \rho}$ satisfy \eqref{e:cell_white}.
Now if we substitute the series expansion in (\ref{e:ef_dif_alpha}) we obtain:
\begin{displaymath}
\alpha \cdot \cK \alpha = \frac{1}{\sqrt{\tau }} \langle \alpha \cdot y \phi_{0}^{\alpha}
\rangle_{\rho^{\eta}\widehat{\rho}(x,y)} +\mathcal{O}(\sqrt{\delta}).
\end{displaymath}
Since $\phi_{0}^{\alpha}$ does not depend on $\eta$ we can integrate over the $\eta$ variable and obtain
\begin{eqnarray}
\alpha \cdot \cK \alpha &=& \frac{1}{\sqrt{\tau }} \langle \alpha \cdot y \phi_{0}^{\alpha}
\rangle_{\widehat{\rho}(x,y)} +\mathcal{O}(\sqrt{\delta}).\\
&=& \alpha \cdot \widehat{\cK} \alpha +\mathcal{O}(\sqrt{\delta})
\end{eqnarray}

\rem \normalfont Thus we have shown that for $\delta \ll 1$ the effective
diffusivity of the colored noise problem  is approximately  equal to  that arising from the   white
noise problem up to terms of $\mathcal{O}(\delta),$ provided the centering condition is
satisfied. It is straightforward to show that exactly the same result holds even when the
centering condition is not satisfied. In this case the asymptotic analysis is
done for  equations \eqref{e:deff_drift} and \eqref{e:cell_drift}; the
effective  drift $V$ vanishes in the limit $\delta \rightarrow 0$.

%
%
%
\section{Numerical Investigations}
\label{sec:numerics}
In this section we study the dependence of the effective diffusivity
\eqref{e:eff_diff_form} or \eqref{e:eff_diff:white}  on the various parameters of the
problem (Stokes number, molecular diffusivity etc.) by means of numerical experiments. We  study equations \eqref{e:motion_resc},\eqref{e:kraichnan} in two dimensions with the
velocity field being the Taylor--Green flow, modulated in time by a one dimensional OU
process. The equations of motion for the colored velocity field are
\begin{equation}\label{e:motion_num_col}
\tau \ddot{x} = \mu \nabla^{\bot} \psi_{TG}(x) - \dot{x} + \sigma \xi,
\quad \dot{\mu} =- \frac{\alpha}{\delta} \mu + \frac{\lambda}{\delta} \zeta,
\end{equation}
with
\begin{equation} \label{e:TG}
\psi_{TG}(x)=\sin{(x_{1})}\sin{(x_{2})}.
\end{equation}
Note that we consider the original equations \eqref{e:the_model},\eqref{e:vel},\eqref{e:ou_model}
rather than the rescaled version \eqref{e:main_model} .The white noise model is
\begin{equation}\label{e:motion_num_white}
\tau \ddot{x} = \alpha^{-1} \lambda \nabla^{\bot} \psi_{TG}(x) \zeta - \dot{x} +
\sigma \xi.
\end{equation}
Here $\zeta, \xi$ are independent Gaussian white noise processes in dimensions
$1$ and $2$ respectively.

Our aim is to study the dependence of the effective diffusivity on the parameters
$\alpha, \lambda$ and $\delta$. The Taylor-Green Flow satisfies the parity invariance
condition \eqref{e:parity} and consequently the centering condition \eqref{e:centering} is satisfied.
Furthermore, the symmetry properties of the Taylor--Green flow imply that the two
diagonal components of the effective diffusivity are equal,
while the off--diagonal components vanish. For the rest of the section we will use the
notation $\mathcal{K}=\mathcal{K}_{11}=\mathcal{K}_{22}$ and will refer to $\cK$ as the
\emph{effective diffusivity}.
\begin{figure}
\begin{center}
\subfigure[$\sigma=0.001$]{\includegraphics[scale=0.38]{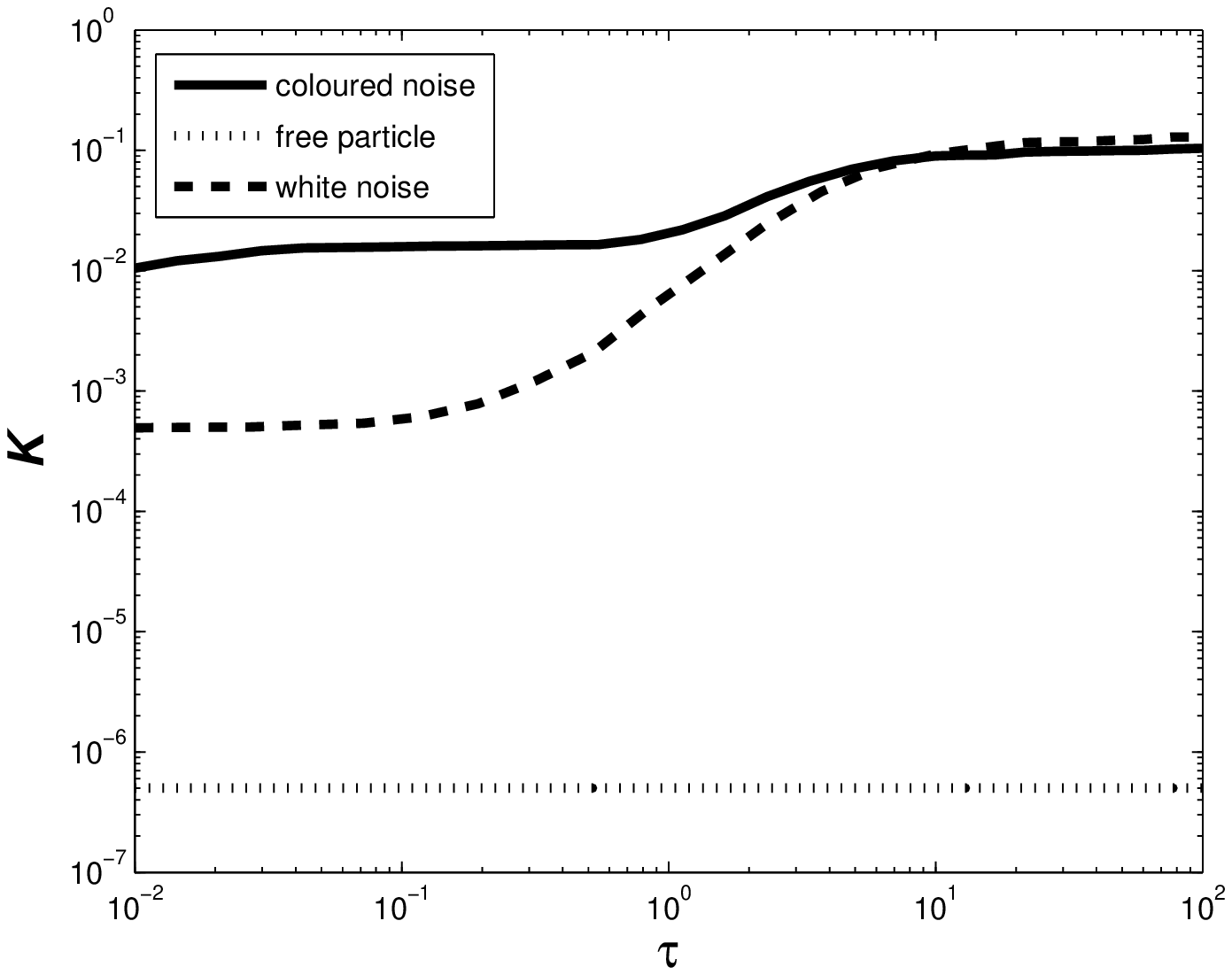}}
\subfigure[$\sigma=0.01$]{\includegraphics[scale=0.38]{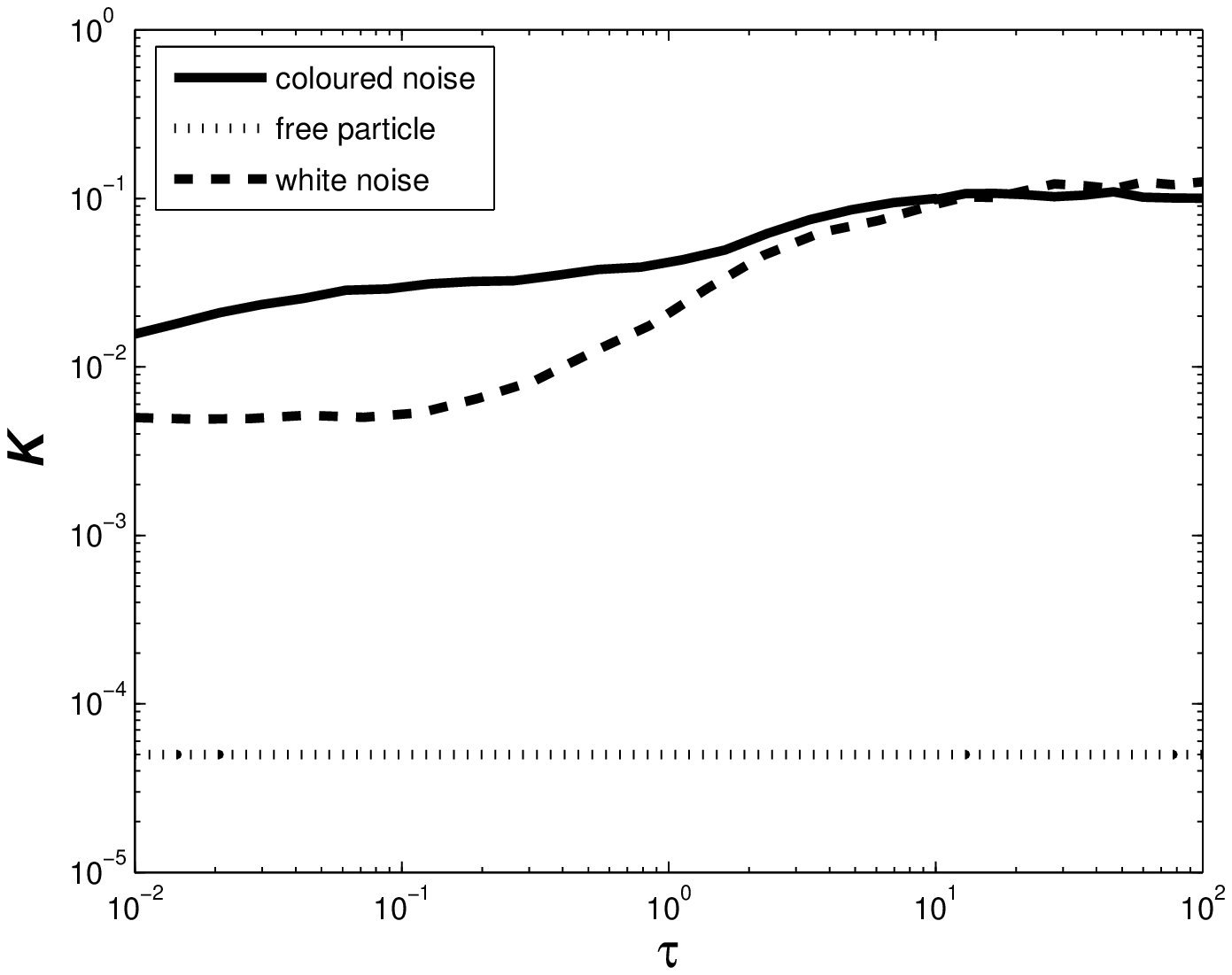}}
\caption{Effective diffusivity as a function of $\tau$ for $\sigma \ll 1$.}
\label{fig:K_tau_sig10_3}
\end{center}
\end{figure}

We calculate the effective diffusivity using Monte Carlo simulations, rather than
solving the Poisson equations \eqref{e:cell_color},\eqref{e:cell_white}. The numerical
solution of degenerate Poisson equations of this form is an interesting problem which we
leave for future study. We solve equations \eqref{e:motion_num_col} and
\eqref{e:motion_num_white} numerically for different realizations of the noise and we
compute the effective diffusivity using its Lagrangian definition
$$
\lim_{t \rightarrow \infty} \frac{1}{2t} \langle (x(t)
    - \langle x(t) \rangle) \otimes (x(t) - \langle x(t) \rangle)
    \rangle=\mathcal{K}I.
$$
where $\langle \cdot \rangle$ denotes ensemble average over all driving Brownian motions.
In practice, of course, we approximate the ensemble average by a finite number of
ensemble members. We solve the equations (\ref{e:motion_num_col}),
(\ref{e:motion_num_white}) using the Euler--Marayama method for the $x$-variables  and
the exact solution for the Ornstein--Uhlenbeck process. The Euler method for the colored
noise  problem  has a order of strong convergence 1 since then noise is additive in this
case\cite{KlPl92} ; in the white noise case this reduces to order $1/2$, since the noise
is then multiplicative. We use 3000 particles  with fixed non random initial conditions.
The initial velocity of the inertial particles is always taken to be $0$. We integrate
over 10000 time units with $\Delta t= 10^{-3}$.

\begin{figure} 
\begin{center}
\subfigure[$\sigma=0.1179$]{\includegraphics[scale=0.38] {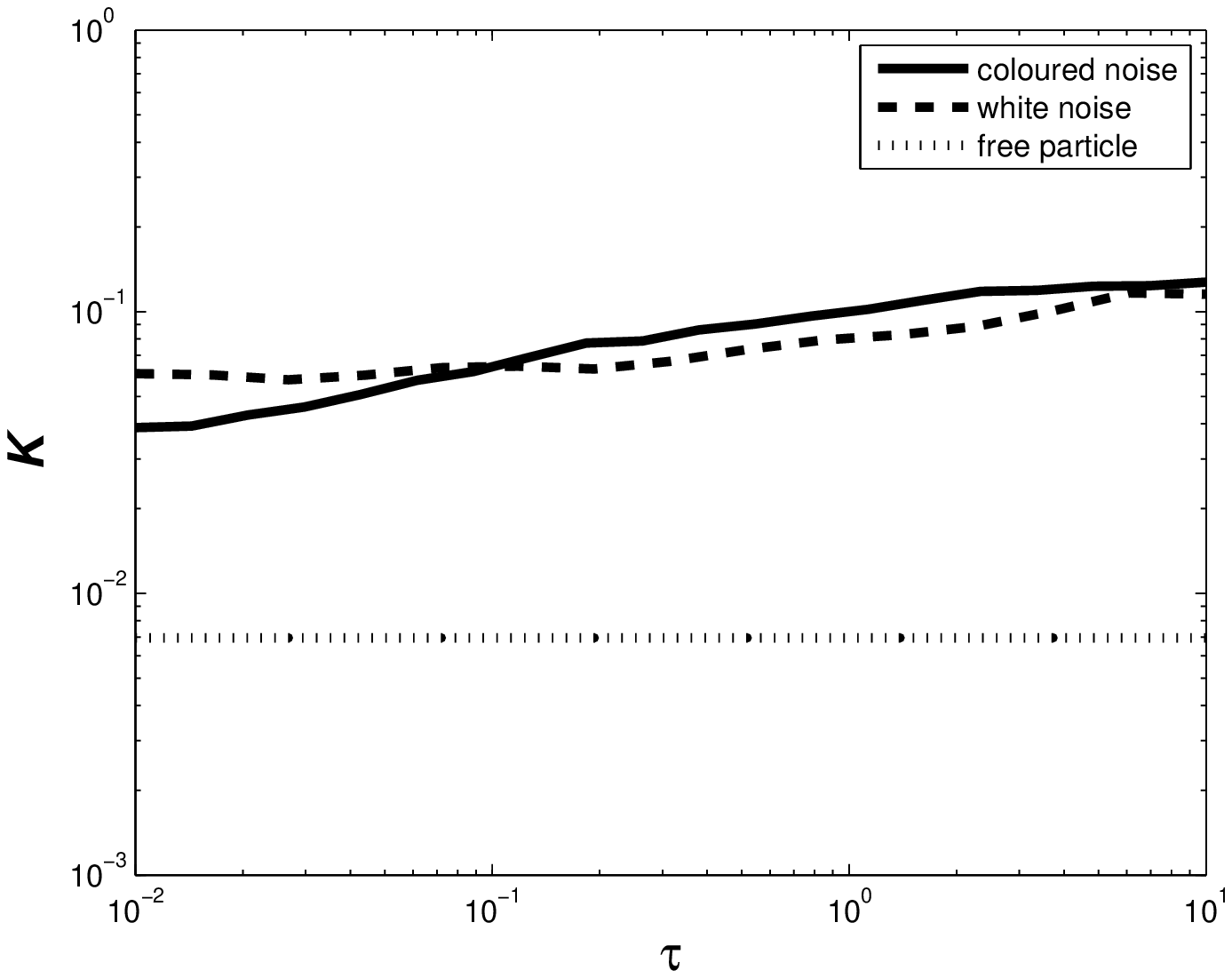}}
\subfigure[$\sigma=1.3895$] {\includegraphics[scale=0.38]{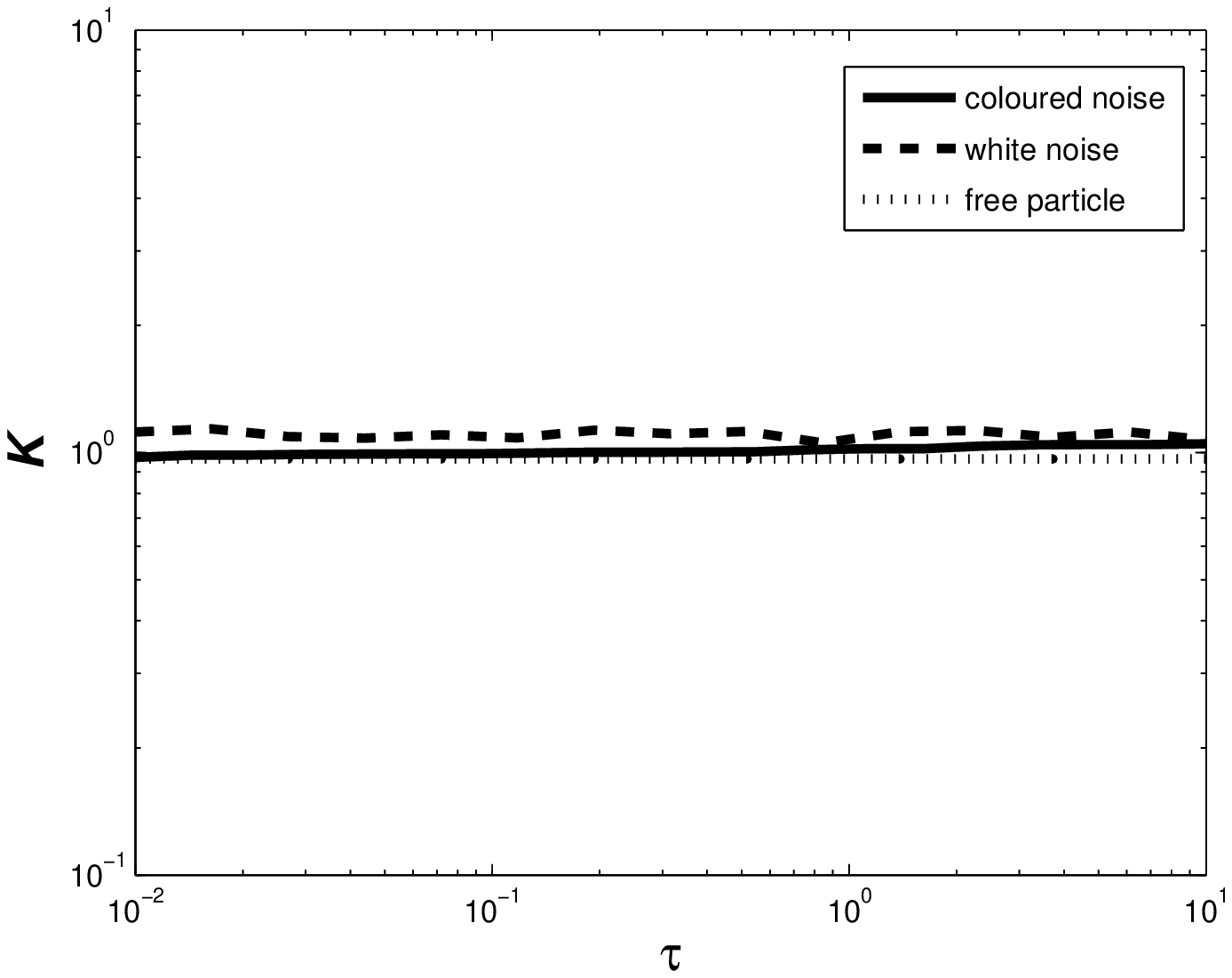}}
\caption{Effective diffusivity as a function of $\tau$ for $\sigma = \cO(1)$.}
\label{fig:K_tau_sig10_2}
\end{center}
\end{figure}
%

%
%
\subsection{The effect of $\tau$ on the diffusivity}
First we investigate the dependence of the effective diffusivity on the Stokes number
$\tau$ for the Taylor-Green flow. We set  the values of $\lambda = \alpha = \delta=1$. Our results are presented in Figures \ref{fig:K_tau_sig10_3} and
\ref{fig:K_tau_sig10_2}. For comparison we also plot the diffusion coefficient of the free
particle $\sigma^{2}/2$.

We observe that when $\sigma \ll 1$ the effective diffusivity is several orders of magnitude
greater than the molecular diffusivity, both for the colored and the white noise case.
Furthermore, the dependence of $\cK$ on $\tau$ is different when $\tau \ll 1$ and
$\tau \gg 1$, with a crossover occuring for $\tau = \cO(1)$. On the other hand, the enhancement in the diffusivity becomes
much less pronounced when $\sigma$ is not very small, and essentially dissapears as $\sigma$ increases, see Figure \ref{fig:K_tau_sig10_2}. This is to be expected, of course.

%
%
\subsection{The effect of $\sigma$ on the diffusivity}
\begin{figure}
\begin{center}
\includegraphics[scale=0.50]{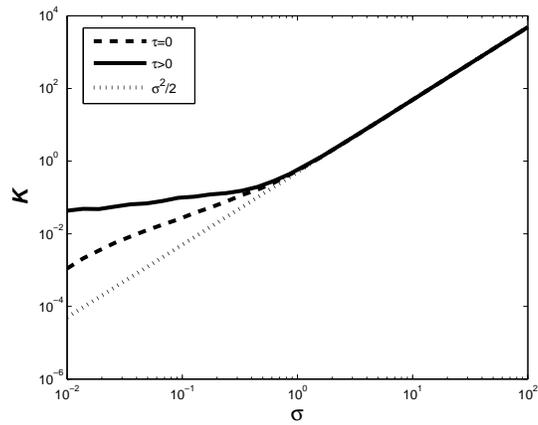}
\caption{Effective diffusivity as a function of $\sigma$  for the colored noise problem}
\label{thal}
\end{center}
\end{figure}

\begin{figure}
\begin{center}
\includegraphics[scale=0.50]{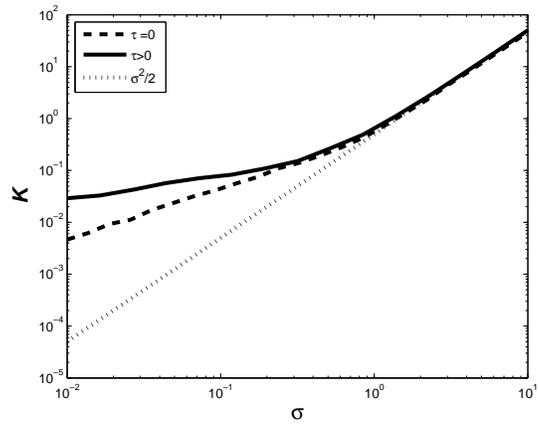}
\caption{Effective diffusivity as a function of $\sigma$ for the white noise problem}
\label{thal1}
\end{center}
\end{figure}

We fix now $\alpha = \lambda = \delta=  1$ and investigate the dependence of $\cK$ on
$\sigma$ for various values of $\tau$. Our results are presented in Figures \ref{thal}
and \ref{thal1}, where for comparison we also plot the diffusion coefficient of the free
particle $\sigma^2/2$ .

In Figure \ref{thal} we plot the effective diffusivity of the colored noise problem in
the case where $\tau=1.0$ (inertial particles) and $\tau=0$ (passive tracers). In both
cases the effective diffusivity is enhanced in comparison with the one of the free
particle problem. However, the existence of inertia enhances further the diffusivity.
This phenomenon has been observed before \cite{PavSt05b} in the case where the velocity
field used was again the Taylor-Green velocity field but with no time dependence.

In Figure \ref{thal1} we plot the effective diffusivity of the white noise problem as a function of
$\sigma$ in the case where   $\tau=1.3895$(inertial particles) and $\tau=0$(passive tracers). The enhancment occurs in both cases but again  the existence of inertia enhances further the diffusivity. As expected when  $\sigma >>1$ the effective diffusivities for both inertial particles and passive tracers converge to $\frac{\sigma^{2}}{2}$.

%
%
%
%
\subsection{The effect of $\alpha$ and $\lambda$ on the effective diffusivity}

\begin{figure} 
\begin{center}
\includegraphics[scale=0.50]{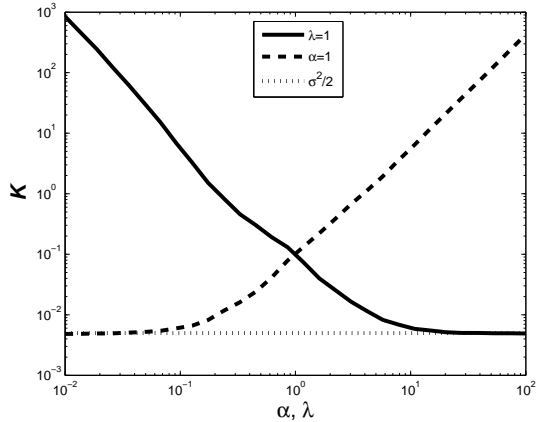}
\caption{Effective diffusivity as a function of $\alpha$, $\lambda$ }
\label{fig:K_vs_alpha_lambda}
\end{center}
\end{figure}

In this subesection we investigate the dependence of $\cK$ on $\alpha$ and $\lambda$ for
$\sigma = 0.1, \, \tau = \delta= 1.0$. In the limit as either $\alpha \rightarrow \infty$ or
$\lambda \rightarrow 0$ the OU processes
converges to $0$. It is expected, therefore, that in either of these two limits the solution
of the Stokes equation converges to the solution of
\begin{equation} \label{e:free}
\tau \ddot{x}=-\dot{x}+\sigma \xi(t),
\end{equation}
and, consequently, in this limit the effective diffusivity is simply the molecular
diffusion coefficient. This result can be derived using techniques from e.g.
(Chapter 9 in \cite{HorsLef84}). On the other hand, when either $\alpha \rightarrow 0$ or $\lambda
\rightarrow \infty$, the OU process dominates the behavior of solutions to the Stokes
equation and, consequently, the effective diffusivity is controlled by the OU process.
The above intuition is supported by the numerical experiments presented in Figure
\ref{fig:K_vs_alpha_lambda}. In particular, the effective diffusivity converges to
$\frac{\sigma^2}{2}$ when either $\alpha$ becomes large or $\lambda$ becomes small, and
becomes unbounded in the opposite limits.
%
%
\subsection{The effect of $\delta$ on the diffusivity}
\begin{figure}
\begin{center}
\includegraphics[scale=0.50]{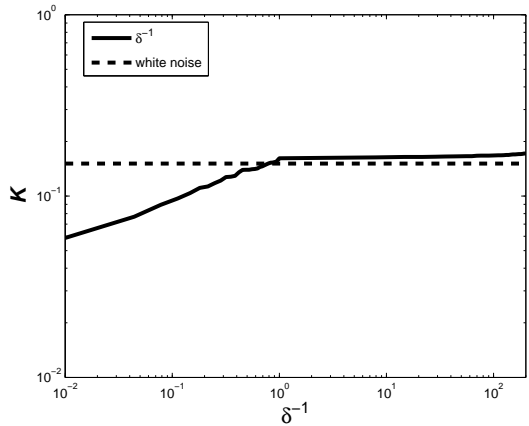}
\caption{Effective diffusivity as a function of $\frac{1}{\delta}$ }
\label{thala}
\end{center}
\end{figure}

In this section we study the effect of $\delta$ in the effective diffusivity of the colored
noise problem. Our results are plotted in Figure \ref{thala}. The values of $\alpha,
\lambda$ are set equal to 1, while $\tau=1.3895$ and $\sigma=0.3162$.

We expect that as $\delta \rightarrow 0$ the colored noise problem should approach the
white noise problem. This is what we see in Figure \ref{thala}, since when $\delta$ is of
$\mathcal{O}(1)$  the value of the effective diffusivity for the colored noise problem is
almost the same as  the white noise one. The rate at which the effective diffusivity for
the colored noise problem converge to the one for the white noise problem depends on the
values of $\tau$, $\sigma$. Indeed, as we have already seen in subsection 5.1 for small
values of $\tau$ and $\sigma$ there is a significant difference between the values for
the two diffusivities when $\delta=\mathcal{O}(1)$.

\section{Conclusions}
The problem of homogenization for inertial particles moving in a time dependent random
velocity field was studied in this paper. It was shown, by means of formal multiscale
expansions as well as rigorous mathematical  analysis, that the long-time, large-scale
behavior of the particles is governed by an effective Brownian motion. The covariance of
the limiting Brownian motion can be expressed in terms of the solution of an appropriate
Poisson equation.

The combined homogenization/rapid decorrelation in time for the velocity field limit was
also studied. It was shown that the two limits commute.

Our theoretical findings were augmented by numerical experiments in which the dependence
of the effective diffusivity on the various parameters of the problem was investigated.
Furthermore, various limits of physical interest--such as $\sigma \rightarrow 0, \, \tau
\rightarrow 0$ etc.--where studied. The results of our numerical experiments suggest that
the effective diffusivity depends on the various parameters of the problem in a very
complicated, highly nontrivial way.

There are still many questions that remain open. We list some of them.
\begin{itemize}
\item Rigorous study of the dependence of the effective diffusivity on the various
parameters of the problem. This problem has been studied quite extensively in the context
of passive tracers. Apart for this being an interesting problem for the point of view of
the physics of the problem, it also leads to some very interesting issues related to the
spectral theory of degenerate, nonsymmetric second order elliptic operators.
\item Numerical experiments for more complicated flows. It is expected that the amount of
enhancement of the diffusivity will depend sensitively on the detailed properties of the
incompressible, time dependent flow.
\item Proof of a homogenization theorem for infinite dimensional OU processes, i.e. for
the model~\eqref{e:motion}. In this setting, questions such as the dependence of the
effective diffusivity on the energy spectrum and the regularity of the flow can be
addressed.
\end{itemize}
%
%
\appendix
\section{Proof of the Homogenization Theorem}
\label{app:proof}
Let $x(t): \R^+ \mapsto \R^d$ be the solution to the SDE
\begin{equation} \label{e:main}
\tau \ddot{x}(t)=u(x(t),t)-\dot{x}(t)+\sigma \dot{\beta}_1(t),
\end{equation}
where $\tau, \, \sigma > 0$, $\beta_1(t)$ is a standard Brownian motion on
$\mathbb{R}^{d}.$  Furthermore the field $v(x,t): \R^d \times \R^+ \mapsto \R^d$ is
given by
$$
u(x,t) = F(x) \mu(t).
$$
Here, for each fixed $x$, $F(x) \in \R^{d \times n}$ and, furthermore,
$F(x)$ is smooth and
period $1$ as a function of $x$.
Also $\mu(t): \R^+ \mapsto \R^n$ is the solution of
\begin{equation} \label{e:ou}
\dot{\mu}(t)=- \delta^{-1} A \mu(t) + \delta^{-1} \sqrt{\Lambda}\dot{\beta}_2(t).
\end{equation}
where $\beta_2(t)$ is a standard Brownian motion on $\R^n$ and $A, \, \Lambda$ are $n
\times n$ positive definite matrices. Our goal is to prove that the rescaled process
\begin{equation}
x^\eps(t) := \eps x \left(\frac{t}{\eps^2} \right)
\end{equation}
converges weakly to a Brownian motion with variance given by \eqref{e:eff_diff_form}.
We rewrite \eqref{e:main}, \eqref{e:ou}
as a system of first order SDEs
\begin{subequations} \label{e:syst}
\begin{eqnarray}
\dot{x} &=&   \frac{1}{\sqrt{\tau}} y, \\
\dot{y} &=& \frac{1}{\sqrt{\tau}}F(x)\mu -\frac{1}{\tau}y+
\frac{\sigma}{\sqrt{\tau}} \dot{\beta_{1}}, \\
\dot{\mu}&=&-\frac{A}{\delta}\mu
 \   + \frac{\sqrt{\Lambda}}{\delta} \dot{\beta_{2}}.
\end{eqnarray}
\end{subequations}
This is a Markov process for $(x(t),y(t),\mu(t))$ on $\IR^{d} \times \IR^{d} \times
\IR^{n}$. We let $z(t)$ denote the function $x(t)/\mathbb{Z}^{d}$ so that $z(t) \in
\mathbb{T}^{d}=\IR^{d}/\mathbb{Z}^{d}$. Since $F$ is $1$-periodic we may view
$(z(t),y(t),\mu(t))$ as a Markov process on $\mathbb{T}^{d}\times \IR^{d} \times
\IR^{n}$.
\begin{theorem}\label{thm:homog}
Let $\{ x(t), \, y(t), \, \mu(t) \}$ be the Markov process defined through the solution
of \eqref{e:syst}, where $A = I$ and $\Lambda = \lambda I, \, \lambda >0$, $\sigma >0$
and assume that the process $\{z(t),y(t),\eta(t)\}$ is stationary. Assume that the vector
field $F(x) \mu$ has zero expectation with  respect to the invariant measure
$\rho(z,y,\mu)dz d y d \mu$ of the Markov process $\{z(t), y(t), \mu(t) \}.$ Then the
rescaled process $x^{\eps}(t)$ converges weakly to a Brownian motion with covariance
matrix $2sym({\cal K})$ where
\begin{equation}\label{e:eff_diff_thm}
\cK = \int_{\T^d \times \R^d \times \R^n} \left(-\cL_0 \Phi \right) \otimes \Phi
\rho(z,y,\mu)dz dy d \mu.
\end{equation}
Here $\Phi(z,y, \mu) \in L^2(\T^d \times \R^d \times \R^n, \rho(z,y,\mu)d z d y d \mu; \R^d)$ is the
unique--up to additive constants--solution of the Poisson equation
\begin{equation}\label{e:poisson}
- \cL_0 \Phi = \frac{1}{\sqrt{\tau}} y.
\end{equation}
\end{theorem}
\begin{remark} \normalfont
The assumptions on $A$ and $\Lambda$ are made merely for notational simplicity. It is
straightforward to extend the proof presented below to the case where $A, \, \Lambda$ are not
diagonal matrices, provided that they are positive definite.
\end{remark}
\begin{remark} \normalfont
In the case where the centering condition \eqref{e:centering}, or equivalently
\eqref{e:centering1}, is not satisfied, then to leading order the particles move
ballistically with an effective velocity $V = \langle F(z) \mu \rangle_\rho$. A central
limit theorem of the form of Theorem \ref{thm:homog} provides us with information on the
fluctuations around the mean deterministic motion. See also Remark \ref{rem:drift}.
\end{remark}
\begin{remark} \normalfont
It is not necessary to assume that the process is started in its stationary distribution
as it will approach this distribution exponentially fast. Indeed, as we prove in
Proposition \ref{prop:inv_meas} below, the fast process is geometrically ergodic. This
implies that for every function $\psi: \T^d \times \R^d \times \R^n \mapsto \IR$ which
does not grow too fast at infinity there exist constants $C, \, \delta$ such that
\begin{equation}\label{e:geom_ergod_estim}
\left| \E \left( \psi(z(t), y(t), \mu(t)) \right) - \int_{\T^d \times \R^d \times \R^n}
\psi(z, y, \mu) \rho(dz dy d \mu) \right| \leq C e^{- \delta t},
\end{equation}
where $\E$ denotes expectation with respect to the law of the process $\{ z(t),\, y(t), \, \eta(t)
 \}$ and $\rho(z,y,\mu)dz dy d \mu$ the unique invariant measure.
We make the stationarity assumption to avoid some technical difficulties.

\end{remark}
As is usually the case with theorems of the form \eqref{thm:homog},e.g. \cite{carmona,
PapStrVar77,komor_olla, landim_olla}, the proof \eqref{thm:homog} is based on the cental
limit theorem for additive functionals of Markov processes: we apply the It\^{o} formula
to the solution of the Poisson equation \eqref{e:cell_color} to decompose the rescaled
process \eqref{e:resc} into a martingale part and a remainder; we then employ the
martingale central limit theorem \cite[Ch. 7]{EthKur86} to prove a central limit theorem
for the martingale part and we show that the remainder becomes negligible in the limit as
$\eps \rightarrow 0$. In order to obtain these two results we need to show that the fast
process is ergodic and that the solution of the Poisson equation \eqref{e:cell_color}
exists and  is unique in an appropriate class of functions, and that it satisfies certain
\emph{a priori} estimates. In order to prove that the fast process is ergodic in a
sufficiently strong sense we use results from the ergodic theory of hypoelliptic
diffusions \cite{MatSt02}. In order to obtain the necessary estimates on the solution of
the Poisson equation \eqref{e:cell_color} we use results on the spectral theory of
hypoelliptic operators \cite{EH00,EH03,EckmPillR-B00,EPR99, HelNi05}. Our overall
approach is similar to the one developed  in \cite{HairPavl04}.

For the proof of the homogenization we will need the following three technical results
which we prove in Appendix~\ref{sec:estim}.
\begin{prop}\label{prop:inv_meas}
Let $\cL_0$ be the operator defined in \eqref{e:oper_defn} and assume that $F(x) \in
C^\infty(\T^d ; \R^n)$ and $\sigma>0$. Then the process $\{z(t), \, y(t), \, \mu(t) \}$
generated by $\cL_0$ is geometrically ergodic.
\end{prop}
\begin{prop}\label{prop:estim_inv_meas}
Assume that $A = I, \, \Lambda = \lambda I, \, \lambda >0$ and $\sigma>0.$ Also let
$\rho(z,y,\mu)$ be the invariant measure of the process generated by $\cL_0$. Then, for
every $\alpha \in (0, 2\sigma^{-2})$ and $\beta \in (0, 2 \lambda^{-1})$ there exists a
function $g(z,y, \mu) \in \cS$ (the Schwartz space of smooth functions with fast decay at
infinity) such that
\begin{equation}\label{e:estim_inv_meas}
\rho(z, y, \mu) = e^{-\frac{\alpha}{2} \|y \|^2 - \frac{\beta}{2} \|\mu \|^2} g(z, y,
\mu).
\end{equation}
\end{prop}
\begin{prop}
\label{prop:boundphi} Let $h \in C^{\infty}(\T^d \times \R^d \times \R^n)$ with
$D_{z,y,\mu}^{\alpha} h \in L^2 (\T^d \times \R^d \times \R^n ; e^{- \epsilon \|y \|^2 -
\eps \|\mu \|^2} dz dy d \mu)$ for every multiindex $\alpha$ and every $\epsilon >0$.
Assume further that $\int h( z, y, \eta)\,\rho(dz\,dy \, d \mu) = 0$, where $\rho$ is the
invariant measure of the process $\{z(t), \ y(t), \ \mu(t) \}$. Then there exists a
solution $f$ of the equation
\begin{equation}\label{e:pois}
- \cL_0 f = h.
\end{equation}
Moreover, for every $\alpha, \, \beta > 0$, the function $f$ satisfies
\begin{equation}\label{e:boundPhi}
f(z,y, \mu) = e^{{\alpha\over 2} \|y\|^2 + {\beta\over 2} \|\eta \|^2} \tilde{f}(z,y,
\mu)\;,\qquad \tilde{f} \in \cS\;.
\end{equation}
Furthermore, for every $\alpha \in (0,2\sigma^{-2}), \, \beta \in (0, 2 \lambda^{-1})$,
$f$ is unique (up to an additive constant) in  $L^2 (\T^d \times \R^d \times \R^n , e^{-
\alpha \|y \|^2 - \beta \|\mu \|^2} dz dy d \mu)$.
\end{prop}
\bigskip

\noindent {\it Proof of Theorem \ref{thm:homog}.} We have already shown that the
centering assumption on the velocity field, equation \eqref{e:centering}, is equivalent
to $\langle y \rangle_\rho = 0$. Moreover, $y$ clearly satisfies the smoothness and fast
decay assumptions of Proposition~ \ref{prop:boundphi}. Proposition \ref{prop:boundphi}
applies to each component of equation \eqref{e:poisson} and we can conclude that there
exists a unique smooth vector valued function $\Phi$ which solves the cell problem and
whose components satisfy estimate \eqref{e:boundPhi}.

We apply now It\^{o} formula to $\Phi(x(t), y(t), \mu(t))$ with $(x,y,\mu)$ solving
\eqref{e:zyeta} and use the fact that $- \cL_0 \Phi = \frac{1}{\tau} y$ to obtain
\begin{eqnarray*}
x^{\eps}(t)  &=& \eps x(t/\eps^{2}) \\
   &=& \eps x(0)+\frac{\eps}{\sqrt{\tau}}\int_{0}^{t/\eps^2}y(s)ds \nonumber \\
     &=& \eps x(0)- \epsilon \left(\Phi(z(t/\eps^2), y(t/\eps^2),\mu(t/\eps^2))-
             \Phi(z(0),y(0),\mu(0))\right))
      \\
         && + \eps \frac{\sigma}{\sqrt{\tau}}\int_{0}^{t/\eps^2}
             \nabla_{y}\Phi(y(s),z(s),\mu(s)) d\beta_{1}(s)
       \\ &&+
         \eps \frac{\sqrt{\lambda}}{\delta} \int_{0}^{t/\eps^2}\nabla_{\mu}\Phi(y(s),z(s),
         \mu(s)) d\beta_{2}(s)
             \nonumber.
       \\ & =: & \eps x(0) + R^\eps_t + M^\eps_t + N^\eps_t.
\end{eqnarray*}
Clearly $\lim_{\eps \rightarrow 0} \eps^{2} \E |x(0)|^2 = 0$. Furthermore, the
stationarity assumption together with Propositions \ref{prop:estim_inv_meas} and
\ref{prop:boundphi} imply that
\begin{eqnarray}
\E |R^\eps_t|^2 \leq C \eps^2 \|\Phi(z,y,\mu) \|^2_{L^2_\rho} \leq C \eps^2.
\end{eqnarray}
Consider now the martingales $M^\eps_t$ and $N^\eps_t$. According to the martingale
central limit theorem \cite[Thm. 7.1.4]{EthKur86}, in order to prove convergence of a
martingale to a Brownian motion, it is enough to prove convergence of its quadratic
variation in $L^1$ to $\sigma^2 \, t$; $\sigma^2$ is the variance of the limiting
Brownian motion. This now follows from propositions \ref{prop:estim_inv_meas} and
\ref{prop:boundphi}, together with the ergodic theorem for additive functionals of
ergodic Markov processes \cite{yor}. In particular, using $\langle \cdot \rangle_t$ to
denote the quadratic variation of a martingale, we have that
\begin{eqnarray*}
\langle M^\eps\rangle_t &=& \eps^2 \frac{\sigma^2}{\tau} \int^{t/\eps^2}_0 \nabla_y
\Phi(x(s),y(s), \mu(s)) \otimes \nabla_y \Phi(x(s),y(s), \mu(s)) \, ds
\\ &
\rightarrow &  \frac{\sigma^2}{\tau} \left\langle \nabla_y \Phi(x,y, \mu) \otimes
\nabla_y \Phi(x,y, \mu) \right\rangle_{\rho} t \quad \mbox{in} \; \; L^1.
\end{eqnarray*}
Similarly
\begin{eqnarray*}
\langle N^\eps \rangle_t \rightarrow   \frac{\lambda}{\delta^2} \left\langle \nabla_{\mu}
\Phi(x,y, \mu) \otimes \nabla_{\mu} \Phi(x,y, \mu)
 \right\rangle_{\rho} t \quad \mbox{in} \; \; L^1.
\end{eqnarray*}
We combine the above with equation~\eqref{e:eff_diff_2} and use the fact that
$\eta=\sqrt{\delta}\mu$ and that $A, \, \Lambda$ are diagonal matrices to conclude the
proof of the theorem. \qed

\begin{remark}\normalfont
With a bit of extra work one can also obtain estimates on the rate of convergence to the
limiting Brownian motion in the Wasserstein metric, as was done in \cite{HairPavl04} for
the case of a time independent velocity field. To accomplish this we need to obtain
appropriate pathwise estimates on the rescaled particle velocity $y(t/\eps^2)$ and the
Ornstein--Uhlenbeck process $\mu(t/\eps^2)$. We also need to introduce an additional
Poisson equation of the type \eqref{e:pois}
and to apply the It\^{o} formula to its solution.
The Poisson equation of type \eqref{e:pois} plays
the role of a higher order cell problem from the theory of homogenization;
see ,e.g.,
\cite{cioran} for the proof of an error estimate using higher order cell problems in the
PDE setting. The argument used in \cite[Thm. 2.1]{HairPavl04} is essentially a pathwise
version of the PDE argument.
We leave the details of this quantative error bound to the interested reader.
\end{remark}
%
%
%
%
\section{Proof of
Propositions~\ref{prop:inv_meas}--~\ref{prop:boundphi}}\label{sec:estim}
In this section we  prove that the operator $\mathcal{L}_{0}$ generates a geometrically
ergodic Markov process. This means that there exists a unique invariant measure $\rho(dz
dy d \mu)$ of the process which has a smooth density $\rho(z, y, \mu)$ with respect to
Lebesgue measure on $\T^d \times \R^d \times \R^n$; and that, furthermore, estimate
\eqref{e:geom_ergod_estim} holds. In addition, we prove some  regularity properties of
the invariant density and existence and uniqueness of solutions together with
 a priori estimates for the Poisson equation \eqref{e:cell_color}.
The proof of Proposition~\ref{prop:inv_meas} follows the lines of \cite{MatSt02}. The
proof of Propositions~\ref{prop:estim_inv_meas} and~\ref{prop:boundphi} is based on
results from~\cite{HairPavl04}.

The proof of this Proposition~\ref{prop:inv_meas} is based upon three lemmas. In the
first lemma we show that the transition probability $P_t$ has a smooth density $\rho_t$
with respect to Lebesgue. In the second we show that  $\rho_t$ is everywhere positive. In
the third we show that there exists a Lyapunov function. These three lemmas imply that
the fast process is geometrically ergodic \cite[Cor. 2.8]{MatSt02}.
\begin{lemma}\label{lem:hormand}
Assume that $F(z) \in C^\infty(\T^d ; \R^n)$. Then the Markov process generated by
$\mathcal{L}_{0}$ has a smooth transition probability density $\rho_t$.
\end{lemma}
\begin{remark}\normalfont
The density $\rho_t$ is the solution of the evolution Fokker--Planck equation
\begin{equation*}
\frac{\partial \rho_t}{\partial t} = \cL_0^* \rho_t.
\end{equation*}
\end{remark}
\proof This follows from Hormander's theorem \cite[Thm. 38.15]{RW00b}.
The Markov process $\{z(t),
\, y(t), \, \mu(t) \}$ generated by $\mathcal{L}_{0}$ solves the SDE
\begin{subequations} \label{e:zyeta}
\begin{eqnarray}
d z& = & y \, dt,\\
d y & = & \frac{1}{\tau}F(z)\mu \, dt - \frac{1}{\tau} y \, dt +
\frac{\sigma}{\tau} \, d W_1, \\
d \mu & =& -A \mu \, dt + \sqrt{\Lambda} \, d W_{2},
\end{eqnarray}
\end{subequations}
with the understanding that $z \in \mathbb{T}^{d}  ,y \in \mathbb{R}^{d}$ and $\mu \in
\mathbb{R}^{n}$.
The Jacobian of the drift for this system is:
\begin{displaymath}
J=\left(\begin{array}{c|c|c}
O_{d \times d} & I_{d \times d} & O_{d \times n}  \\
 \hline
\frac{1}{\tau}DF_{d \times d} & -\frac{1}{\tau}I_{d \times d} & \frac{1}{\tau}F_{d \times n} \\
\hline
O_{n \times d} & O_{n \times d} & -A_{n \times n}
\end{array} \right)
\end{displaymath}
where
\begin{displaymath}
(DF)_{k,l}= \frac{\partial}{\partial x_{l}} \sum_{i=1}^{m} F_{ki}(x)\mu_{i}.
\end{displaymath}
In order to prove that our system is hypo-elliptic we need to span $\IR^{2d+n}$ through the noise vectors and their Lie commutators with the drift. We are going to
study two cases. The first is when $\sigma \neq 0$. In this case the noise provides the vectors:
\begin{displaymath}
e^{(i)}_{j}=\delta_{ij},\ \ \  i= d+1, \ldots 2d+n.\  \text{and} \ j=1, \ldots, 2d+n.
\end{displaymath}
thus we are missing $d$ vectors in order to span $\mathbb{R}^{2d+n}$. We can see that we can
obtain the missing vectors
$e^{(i)}_{j},\ i=1, \ldots d$ in the following way:
\begin{displaymath}
Je^{(i)}=e^{(i-d)}-\frac{1}{\tau} e^{(i)}, \ \ i= d+1, \ldots, 2d
\end{displaymath}
thus we obtained the vectors
$e^{(i)},\  i= 1 \ldots 2d+n$ that span $\mathbb{R}^{2d+n}\qed.$

\begin{remark}\normalfont
We briefly remark on the case where $\sigma=0$. In this case since there is no noise in the
equation describing the motion of $x$ and $y$ we need to span $\mathbb{R}^{2d+n}$ having the
following vectors:
\begin{displaymath}
e^{(i)}_{j}=\delta_{ij},\ \ \  i= 2d+1, \ldots 2d+n.\  \text{and} \ j=1, \ldots, 2d+n.
\end{displaymath}
We  notice that:
\begin{displaymath}
(Je^{(i)})_{j} =
\begin{cases}
0 &  \text{if $j=1, \ldots, d$} \\
F_{(j-2d)(i-2d)} & \text{if $j=d+1, \ldots, 2d$} \\
A_{(j-2d)(i-2d)} \text{if $j=2d+1, \ldots, j=2d+n $}
\end{cases}
\end{displaymath}
So the only way not to obtain the vectors $e^{(i)},\  i= d+1, \ldots, 2d$ is if the following
equation holds:
\begin{displaymath}
F_{lk}(z)=0, \  \forall l \in \{1, \ldots  n\} \  \text{and} \  k \in \{1, \ldots k\}
\end{displaymath}
If this equation does not hold we can obtain the rest of the vectors in the exact same way as we
did in the case where $\sigma \neq 0.$
\end{remark}

\begin{lemma} For all $Z:=(z,y,\mu)\in \mathbb{T}^{d}\times \IR^{d}\times \IR^{n}, \, t>0$ and
open  $\mathcal{O}\subset \mathbb{T}^{d}\times \IR^{d}\times \IR^{n}$, the transition kernel
corresponding to the Markov process $\{z(t), \, x(t), \mu(t) \}$ defined in \eqref{e:zyeta}
 satisfies $P_{t}(z,\mathcal{O}) > 0$.
\end{lemma}

\proof The proof of this result is based on a controllability argument
\cite{StrVar72,EckmPillR-B00, MatSt02}. We start by writing  \eqref{e:zyeta} compactly
in the form
\begin{equation}\label{e:zyeta_2}
d Z = Y(Z) \, dt + \Sigma d W
\end{equation}
with $Y(Z) = [y, \, \frac{1}{\tau} F(z) \mu - \frac{1}{\tau} y, - A \mu]^T$, $W(t) = [W_1(t), \,
W_2(t)]^T$ and
\begin{eqnarray*}
\Sigma=\left(\begin{array}{c c}
O_{d \times d} & O_{d \times n}  \\
\frac{\sigma}{\tau} I_{d \times d} & O_{d \times n} \\
O_{d \times d} & \sqrt{\lambda}
\end{array} \right).
\end{eqnarray*}
The control problem associated with \eqref{e:zyeta} is
\begin{equation} \label{e:control}
\frac{dR}{dt}=Y(R)+\Sigma\frac{dU}{dt}.
\end{equation}
For any $t>0$, any $a \in \mathbb{T}^{d}\times\IR^{d}\times\IR^{n}$, and any $a^{+} \in
\mathbb{T}^{d}\times\IR^{d}\times\IR^{n}$ we can find smooth $U \in C^{1}([0,t],\IR^{d+n})$
such that \eqref{e:control} is satisfied and $X(0)=a,X(t)=a^{+}$. To see this set $R=((X^{T},
\frac{dX}{dt}^{T},M^{T})^{T})$. The equation for $M$ is
\begin{displaymath}
\frac{dM}{dt}=-AM+\sqrt{\Lambda}\frac{dU_{2}}{dt}.
\end{displaymath}
We consider this equation separately since it does not involve any of the other state
variables. Choose $M$ to be a $C^{\infty}$ path such that, for the given $t>0$,
\begin{equation} \label{e:control_orbit}
M(0)=a_{3}, M(t)=a^{+}_{3}
\end{equation}
where with $a_{3}, a_{3}^{+}$ we denote the last $n$ components of the  vectors  $a,a^{+}$ .
Since $\sqrt{\Lambda}$ is positive definite is invertible and $\frac{dU_{2}}{dt}$ is defined by
substitution and will be as smooth as $M$ and $(\sqrt{\Lambda})^{-1}$ hence $C^{\infty}$. Also
$U(0)$ can be taken as 0.
 The equation for $(X^{T},\frac{dX}{dt}^{T})^{T})$ is
\begin{displaymath}
\tau \frac{d^{2}X}{dt^{2}}+ \frac{dX}{dt}+F(X)M=\sigma \frac{dU_{1}}{dt}.
\end{displaymath}
where $U_{2}$ is chosen that $M$ satisfies \eqref{e:control} .  Now let X be a $C^{\infty}$
path such that, for that given t
\[
\left( \begin{array}{c} X(0) \\
                \frac{dX}{dt}(0)
                \end{array}
\right) = a_{1}
\hspace{10mm}
\left( \begin{array}{c} X(t) \\
                \frac{dX}{dt}(t)
                \end{array}
\right) = a^{+}_{1}
\]
where with $a_{1}, a^{+}_{1}$ we denote the first $2d$ elements of the vectors $a,a^{+}$
respectively. Since $\sigma$ is everywhere invertible,  $\frac{dU_{1}}{dt}$ is defined by
substitution and will be as smooth as $F(X)M$ and
$\sigma^{-1}$ -- hence $C^{\infty}$. Also $U_{1}(0)$ can be taken as 0.

Now  note that the event
\begin{displaymath}
\sup_{0 \leq s \leq t}||W_{2}(s)-U_{2}(s)|| \leq \epsilon
\end{displaymath}
occurs with positive probability since the Wiener measure of any tube is positive. So the
 Brownian motion $\beta_{2}$ controls the $\eta$ component of equation (\ref{e:zyeta}).
Again note that the event
\begin{displaymath}
\sup_{0 \leq s \leq t}||W_{1}(s)-U_{1}(s)|| \leq \epsilon
\end{displaymath}
occurs with positive probability since the Wiener measure of any such tube is positive.
Combining now these two results  is possible to deduce the required open set irreducibility. \qed

\begin{lemma}
Let $\lambda_{1}$ be the smallest eigenvalue of $A$ and $F=\sup_{x \in \mathbb{T}^{d}}||F(x)||$. Then
there exists a constant $\beta>0$ such that the function $V(x,y,\mu)=1+
\frac{2\tau}{2}||y||^{2}+\frac{\tau^{2}F^2+1}{2\lambda_1}||\mu||^{2}$
satisfies
\[
\mathcal{L}_{0}\bigl(V(x,y,\mu)\bigr) \leq -V(x,y,\mu)+\beta
\]
\end{lemma}

\proof
We have that $V(x,y,\mu)$ maps the state space onto $[1,\infty)$ and that
$\lim_{||y||,||\mu||\rightarrow \infty}=+\infty$. Moreover we have
\begin{eqnarray*}
\mathcal{L}_{0}\bigl(V(x,y,\mu)\bigr)&=& F(x)\mu \cdot
2\tau y-2||y||^{2}+\frac{\sigma^{2}}{2}d-\frac{\tau^{2}F^2+1}{\lambda_1}A\mu \cdot
\mu +\frac{trace(\Lambda)}{2}\nonumber \\
&\leq& \tau^{2}F^{2} ||\mu||^{2}+\frac{4\tau^{2}}{4\tau^{2}}||y||^{2}-2||y||^{2}-
\frac{(\tau^{2}F^2+1)\lambda_{1}}{\lambda_{1}}||\mu||^{2}\\
&+&\frac{trace(\Lambda)}{2} +\frac{\sigma^{2}}{2}d  \nonumber \\
&\leq&-V(x,y,\mu)+\beta, \nonumber
\end{eqnarray*}
with $\beta=\frac{\sigma^{2}d}{2}+\frac{trace(\Lambda)}{2}+1$. \qed
%

\bigskip

{\it Proof of Proposition \ref{prop:inv_meas}.} It follows from the above three lemmas
and \cite[Cor. 2.8]{MatSt02}. \qed

\bigskip

Using results from \cite{EPR99, EH00} we can also derive some regularity estimates for
the invariant density. In addition, we can show that the operator $\cL_0^*$, the formal
$L^2$--adjoint of $\cL_0$, has compact resolvent and, hence, Fredholm theory applies.

\bigskip

\noindent {\it Proof of Proposition \ref{prop:estim_inv_meas}} The proof of this result
is similar to the proof of \cite[Thm. 3.1]{HairPavl04}, which in turn follows the lines
of \cite{EPR99,EH00}. Denote by $\phi_t$ the (random) flow generated by the solutions to
\eqref{e:zyeta} and by $\CP_t$ the semigroup defined on finite measures by
\begin{equation}
\left(\CP_t\mu\right)(A) = \E \left(\mu \circ \phi_t^{-1}\right)(A)\;.
\end{equation}
By Lemma \ref{lem:hormand} $\CP_t$ maps every measure into a measure with a
smooth density with respect to the Lebesgue measure. It can therefore be restricted
to a positivity preserving contraction semigroup on $L^1(\T^d \times
\R^d \times \R^n,dz\,dy\,d \mu)$. The generator $\LL_0^*$ of $\CP_t$ is the formal
$L^2$--adjoint of $\LL_0$.

We now define an operator $K$ on $L^2(\T^d \times \R^d \times \R^n,dz\,dy \, d \mu)$ by closing
the operator defined on $\CC_0^\infty$ by
\begin{eqnarray*}
K & = & -e^{\frac{\alpha}{2} \|y\|^{2} + \frac{\beta}{2} \|\mu \|^2 } \LL^*
e^{- \frac{\alpha}{2} \|y\|^{2} - \frac{\beta}{2} \|\mu \|^2 } \\ 
& = &
 -{\sigma^2 \over 2} \Delta_y - {\lambda \over 2} \Delta_\mu + \beta \left( 1 - \frac{\lambda
 \beta}{2}  \right) \| \mu \|^2 + \alpha \left( 1 - \frac{\sigma^2 \alpha}{2} \right) \| y \|^2
 \\ && + (\sigma^2 \alpha^2 -1) \left( y
 \cdot \nabla_y + \frac{n}{2} \right)  + (\lambda \beta - 1)
 \left( \mu \cdot \nabla_\mu + \frac{d}{2} \right)\\
 & -& \alpha y
 \cdot  F(z, \mu) - \frac{n}{2} - \frac{d}{2}.
\end{eqnarray*}
Note at this point that $\alpha < 2\sigma^{-2}$ and $\beta < 2 \lambda^{-1}$ is required
to make the coefficients of $\|y\|^2$ and $\|\mu \|^2$, respectively, strictly positive.

We can rewrite the above expression in H\"ormander's
``sum of squares'' form as
\begin{equation}
K = \sum_{i=1}^{2d + 2n} X_i^* X_i + X_0\;,
\end{equation}
with
\begin{eqnarray*}
X_i &=& \frac{\sigma}{\sqrt{2}} \partial_{y_i}  \quad \mbox{for $i=1\ldots d$,} \\
X_i &=& \sqrt{\frac{\lambda}{2}} \partial_{\mu_{i-d}} \quad \mbox{for $i = n+1 \ldots (n+d)$} \\
X_i &=& \sqrt{\alpha \left((1-{\alpha \sigma^2 \over 2}\right))}y_{i-n-d}
\quad \mbox{for  $i=(n+d +1)\ldots 2n +d$,} \\
X_i &=& \sqrt{\beta \left((1-{\lambda \beta \over 2}\right))}\mu_{i-2n-d}
\quad \mbox{for   $i=(2n+d +1)\ldots (2n +2d)$,} \\
X_0 &=& (\sigma^2 \alpha^2 -1) \left( y
 \cdot \nabla_y + \frac{n}{2} \right)  + (\lambda \beta - 1)
 \left( \mu \cdot \nabla_\mu + \frac{d}{2} \right) - \alpha y
 \cdot  F(z, \mu) - \frac{n}{2} - \frac{d}{2}.
\end{eqnarray*}
Since $F$ is $C^\infty$ on the torus, it can be checked that the assumptions of
\cite[Thm.~5.5]{EH00} are satisfied with $\Lambda^2 = 1 - \Delta_z - \Delta_y
-\Delta_\mu + \|y\|^2 + \|\mu \|^2$. Combining this with \cite[Lem.~5.6]{EH00}, we see
that there exists $\alpha > 0$ such that, for every $\gamma > 0$,
 there exists a positive constant $C$ such that
\begin{equation}\label{e:boundK}
\|\Lambda^{\alpha+\gamma} f\| \le C \left(\|\Lambda^{\gamma}Kf\| +
\|\Lambda^{\gamma}f\|\right)\;,
\end{equation}
holds for every $f$ in the Schwartz space. Clearly, the operator $\Lambda^2$ has compact
resolvent. This, together with \eqref{e:boundK} with $\gamma = 0$ and
\cite[Prop.~5.9]{EH00} imply that $K$ has compact resolvent.

Notice now that
\begin{equation*}
K^*  =  -e^{-\frac{\alpha}{2} \|y\|^{2} - \frac{\beta}{2} \|\mu \|^2 } \LL
e^{ \frac{\alpha}{2} \|y\|^{2} + \frac{\beta}{2} \|\mu \|^2 }.
\end{equation*}
Thus, $e^{-\frac{\alpha}{2} \|y\|^{2} - \frac{\beta}{2} \|\mu \|^2 }$ is the solution of
the homogeneous equation
\begin{equation}\label{e:Kstar}
K^* e^{-\frac{\alpha}{2} \|y\|^{2} - \frac{\beta}{2} \|\mu \|^2 } = 0.
\end{equation}
The compactness of the resolvent of $K$ implies that there exists a function $g$ such that
$$
K g = 0.
$$
Estimate \eqref{e:boundK}, together with a simple approximation argument imply that
$\|\Lambda^\gamma g\| <\infty$ for every $\gamma > 0$, and therefore $g$ belongs to the
Schwartz space. Furthermore, an argument given for example in \cite[Prop~3.6]{EPR99}
shows that $g$ must be positive. Since one has furthermore
\begin{equation}
\LL^* e^{-\frac{\alpha}{2} \|y\|^{2} - \frac{\beta}{2} \|\mu \|^2 } g = 0\;,
\end{equation}
the invariant density $\rho$ satisfies estimate \eqref{e:estim_inv_meas}. \qed

\bigskip

The ergodicity of the fast process, together with the above proposition enable us to
prove the following  lemma.
\begin{lemma}\label{lem:kernel}
Let $\alpha \in (0,2\sigma^{-2}), \, \beta \in (0, 2 \lambda^{-1})$ and let $K$ be as in
the proof of Proposition \ref{prop:estim_inv_meas}. Then, the kernel of $K$ is
one-dimensional.
\end{lemma}

\proof Let $\tilde g \in \ker K$. Then, by the same arguments as above, $e^{-{\alpha
\over 2}\|y\|^2 - {\beta \over 2} \|\mu \|^2} \tilde g$ is the density of an invariant
signed measure for $\CP_t$. The ergodicity of $\CP_t$ immediately implies $\tilde g
\propto g$. \qed

\bigskip

Now we are ready to prove estimates on the solution of the Poisson equation
\eqref{e:poisson}.

\bigskip

{\it Proof of Proposition~\ref{prop:boundphi}} By hypoellipticity, if there exists a
distribution $f$ such that \eqref{e:pois} holds, then $f$ is actually a $C^\infty$
function.

We start with the proof of existence. Fix $\alpha \in (0,2\sigma^{-2}), \, \beta \in (0,
2 \lambda^{-1})$, consider the operator $K^*$ defined in \eqref{e:Kstar}, and define the
function
\begin{equation*}
u(z,y, \mu) = h(z,y, \eta)\, e^{-{\alpha \over 2}\|y\|^2 - {\beta \over 2} \|\mu
\|^2}\;.
\end{equation*}
It is clear that if there exists $\tilde{f}$ such that $K^*\tilde{f} = u$, then $f =
e^{{\alpha \over 2}\|y\|^2 + {\beta \over 2} \|\mu \|^2}\tilde{f}$ is a solution to
\eqref{e:pois}. Consider the operator $K^* K$. By the considerations in the proof of
Proposition~\ref{prop:estim_inv_meas}, $K^* K$ has compact resolvent. Furthermore, the
kernel of $K^*K$ is equal to the kernel of $K$, which in turn by Lemma~\ref{lem:kernel}
is equal to the span of $g$. Define $\CH = \langle g \rangle_\rho^\bot$ and define $M$ to
be the restriction of $K^* K$ to $\CH$. Since $K^*K$ has compact resolvent, it has a
spectral gap and so $M$ is invertible. Furthermore,
we have that $f \in \CH$, therefore $\tilde{f} = K M^{-1} u$ solves $K^*\tilde{f} = u$
and thus leads to a solution to \eqref{e:pois}.

Since $K^*$ satisfies a similar bound to \eqref{e:boundK} and since $\|\Lambda^\gamma
u\|<\infty$ for every $\gamma > 0$, the bound \eqref{e:boundPhi} follows as in
Proposition~\ref{prop:estim_inv_meas}. The uniqueness of $u$ in the class of functions
under consideration follows immediately from Lemma~\ref{lem:kernel}.  \qed

\begin{remark}\normalfont
Note that the solution $f$ of \eqref{e:pois} may not be unique if we allow for
functions that grow faster than $e^{\sigma^{-2}\|y\|^2 + \lambda^{-1} \|\mu\|^2}$.
\end{remark}
\def\cprime{$'$} \def\cprime{$'$} \def\cprime{$'$} \def\cprime{$'$}
  \def\cprime{$'$} \def\cprime{$'$} \def\cprime{$'$}
  \def\Rom#1{\uppercase\expandafter{\romannumeral #1}}\def\u#1{{\accent"15
  #1}}\def\Rom#1{\uppercase\expandafter{\romannumeral #1}}\def\u#1{{\accent"15
  #1}}\def\cprime{$'$} \def\cprime{$'$} \def\cprime{$'$} \def\cprime{$'$}
  \def\cprime{$'$}

\bibliographystyle{plain}

\end{document}